\documentclass{aa}  

\def\dg{$^{\circ}$}

\def\rsun{R$_\odot$}
\def\msun{M$_\odot$}
\def\var{OGLE-LMC-DPV-062}

\usepackage{graphicx}
\usepackage{tabularx}
\usepackage{txfonts}
\usepackage[hidelinks]{hyperref}
\usepackage{csquotes}
\begin{document}

  \title{Cyclic light variations and accretion disk evolution in the LMC eclipsing binary OGLE-LMC-DPV-062}
   \author{R.E.\,Mennickent
          \inst{1}
                   \and
          G.\,Djura\v{s}evi\'c
     \inst{2,3}
                     \and 
             J.A.\,Rosales
             \inst{1,4} 
               \and
          J.\,Garc\'es
          \inst{1}
    \and 
             J.\,Petrovi\'c
             \inst{2} 
   \and    
             D.R. G. \,Schleicher\inst{5}
          \and
          I.\,Soszy{\'n}ski
          \inst{6}
          }
   \institute{Universidad de Concepci\'on, Departamento de Astronom\'{\i}a, Casilla 160-C, Concepci\'on, Chile\\
              \email{rmennick@udec.cl}
         \and
             Astronomical Observatory, Volgina 7, 11060 Belgrade 38, Serbia
                     \and
                     Issac Newton Institute of Chile, Yugoslavia Branch, 11060, Belgrade, Serbia       
                      \and
                    Instituto de F\'isica y Astronom\'ia, Universidad de Valpara\'iso, Chile 
                     \and
                     Dipartimento di Fisica, Sapienza Università di Roma, Piazzale Aldo Moro 5, 00185 Rome, Italy
                     \and
                     Astronomical Observatory, University of Warsaw, Al. Ujazdowskie 4, 00-478 Warszawa, Poland
             }

\authorrunning{Mennickent, Djura\v{s}evi\'c, Rosales et al.}
\titlerunning{The binary OGLE-LMC-DPV-062}

  \abstract
   { Many intermediate-mass close binaries exhibit photometric cycles longer than their orbital periods, likely linked to variations in their accretion disks. Previous studies suggest that analyzing historical light curves provides key insights into disk evolution and may help track changes in mass transfer rates in such systems.}
   {Our research explores short- and long-term fluctuations in the eclipsing system \var, focusing on the variability of its long cycle. We aim to clarify the role of the accretion disk in these modulations, especially those spanning hundreds of days, and to determine the system's evolutionary status to better understand its stellar components.}
   {We analyzed 32.3 years of photometric time series from the Optical Gravitational Lensing Experiment (OGLE) in $I$ and $V$ bands, and from the MAssive Compact Halo Objects (MACHO) project in $B_{M}$ and $R_{M}$ bands. Using data from multiple epochs, we model the accretion disk across 20 equally spaced phases of the long cycle. To solve the inverse problem, we implemented an optimized simplex algorithm to determine the best parameters for the stars, their orbit, and the disk. The Modules for Experiments in Stellar Astrophysics (\texttt{MESA}) code was employed to assess the system's evolutionary stage and predict its past and future development. } 
   { We find an orbital period of 6\fd904858(15) and a long cycle of 229\fd7. Our orbital solutions reproduce the light curves, but the quasi-conservative mass transfer scenario yields rates too high for the orbital period stability. We find consistency with the observed orbital-to-long-period ratio under the magnetic dynamo hypothesis.
The normalized mass transfer rate follows the long cycle, reaching a maximum when brightness is minimum. At that phase, the disk's inner edge thickens, obscuring more of the gainer star. Disk variability mainly occurs  in its vertical extension, with a standard deviation of 69\% the mean value at the inner border, with minor changes in outer radius and temperature;  7\% and 5\% respectively.} {}
\maketitle

\section{Introduction}

Interacting binary stars serve as intricate natural laboratories for exploring a wide array of astrophysical processes. These processes include accretion disk physics, stellar winds, gas dynamics in mass transfer, angular momentum loss and redistribution, stellar rotation, and tidal interactions. A substantial fraction of stars are thought to belong to multiple systems, with close gravitational interactions between stellar components being a common occurrence across the Universe. Some of the most energetic events observed to date -- such as black hole or neutron star mergers -- are interpreted as the end products of binary evolution involving previous episodes of mass exchange and angular momentum loss.  The study of interacting close binaries is rooted in the foundational framework established by \citet{1971ARA&A...9..183P}, who described the key evolutionary processes governing mass exchange in such systems. The geometrical constraints imposed by Roche-lobe overflow are commonly quantified using the widely adopted approximation of \citet{1983ApJ...268..368E}, while the interpretation of eclipsing-binary light curves has long relied on the formalism introduced by \citet{1971ApJ...166..605W}. Together, these works provide the basic theoretical and methodological context for analyzing semi-detached interacting binaries. A comprehensive review of binary star evolution can be found in \citet{2011epbm.book.....E}.

Among the diverse family of interacting binary stars, a particularly intriguing subclass -- known as Double Periodic Variables (DPVs) -- exhibits two distinct photometric periodicities: an orbital  period and a longer, quasi-cyclic variability whose origin remains unresolved \citep{Mennickent2003, Mennickent2017, 2025A&A...701A..90G}. These long cycles typically exceed the orbital period by a factor of 20--40, with amplitudes around 0.1--0.2 mag in the $I$ band. DPVs are semi-detached binaries comprising a B-type main-sequence star (the gainer) surrounded by an optically thick accretion disk, which is sustained by mass transfer from a less massive, Roche-lobe-filling late-type giant star (the donor) with a typical mass of around one solar mass. The B-type component has been "rejuvenated" by the accretion of hydrogen-rich material from its companion \citep{2024A&A...689A.154R}. 

More than 200 DPVs have been identified in the Milky Way and the Magellanic Clouds \citep{Mennickent2003, 2010AcA....60..179P, Pawlak2013, Mennickent2016, 2021ApJ...922...30R, 2024AcA....74..241G}. Detailed studies show that these systems tend to exhibit higher luminosities and effective temperatures than typical Algol-type binaries, likely reflecting different evolutionary states or mass transfer histories \citep{Mennickent2016}. Notable Galactic examples of DPVs include RX\,Cas \citep{1944ApJ...100..230G}, AU\,Mon \citep{1980A&A....85..342L}, $\beta$ Lyrae \citep{1989SSRv...50...35G}, V\,360 Lac \citep{1997A&A...324..965H}, and CX\,Dra \citep{1998HvaOB..22...17K}.

The nature of the long photometric cycle remains an open question. One of the most promising hypotheses invokes a magnetic dynamo operating within the convective envelope of the donor star \citep{2017A&A...602A.109S}. This mechanism could modulate the star's quadrupole moment, producing periodic variations in the mass transfer rate and, consequently, in the structure and brightness of the accretion disk. In this scenario, the observed photometric changes during the long cycle arise from alterations in the disk's radial and vertical dimensions. While this model is supported by correlations between the long-cycle phase and disk properties in several systems \citep[e.g.][]{2018MNRAS.477L..11G, 2025A&A...693A.217M, 2025A&A...698A..56M}, direct evidence of such a stellar dynamo is still lacking. Further observational and theoretical studies are required to confirm its existence and fully characterize the long-term variability of DPVs.

In this paper we present the first combined photometric analysis of \var, covering 30 years of data, to study the disk variability along the long cycle and its evolutionary implications. This binary is also known as OGLE-LMC-ECL-12848, OGLE J051941.10-693117.1, MACHO 78.6343.81  \citep{2011AcA....61..103G, 2016AcA....66..421P}. 
 The astrometric and photometric properties of the system include $\alpha_{2000}$=05:19:41.00, $\delta_{2000}$=$-$69:31:17.0, $B$= 15.960 mag, $V$= 15.925 mag, $R$= 15.863 mag and $I$= 15.823 mag and $V$= 18.001 mag\footnote{https://simbad.cfa.harvard.edu/simbad/}.  The Gaia DR3 catalogue provides 
mean magnitudes G = 15.9879 $\pm$ 0.0095 mag, BP = 15.9160 $\pm$ 0.0238, RP = 15.7616 $\pm$ 0.0228 mag and BP-RP = 0.1544 $\pm$ 0.0329 mag\footnote{https://gea.esac.esa.int/archive/}. The same source provides  extinction $A_G$ = 0.6391 (0.6317,0.6468) mag. The target was classified as a DPV with orbital period 6\fd9044 $\pm$ 0\fd0010 and a long period 226 $\pm$ 13 days \citep{Mennickent2003}.  Later,  
the reported orbital period was improved to 6\fd904830 $\pm$ 0\fd000015 and the long period to 229\fd080 $\pm$ 0\fd062 \citep{2010AcA....60..179P}.

\var\, was selected for this study because of its relatively large photometric variability produced by the combined orbital and long-term cycles. In addition, 
its long-term modulation is traced across more than three decades of photometry, enabling a quantitative test of the hypothesis that changes in the disk structure produce the long-term cycle.

\begin{figure*}
\scalebox{1}[1]{\includegraphics[angle=0,width=18.5cm]{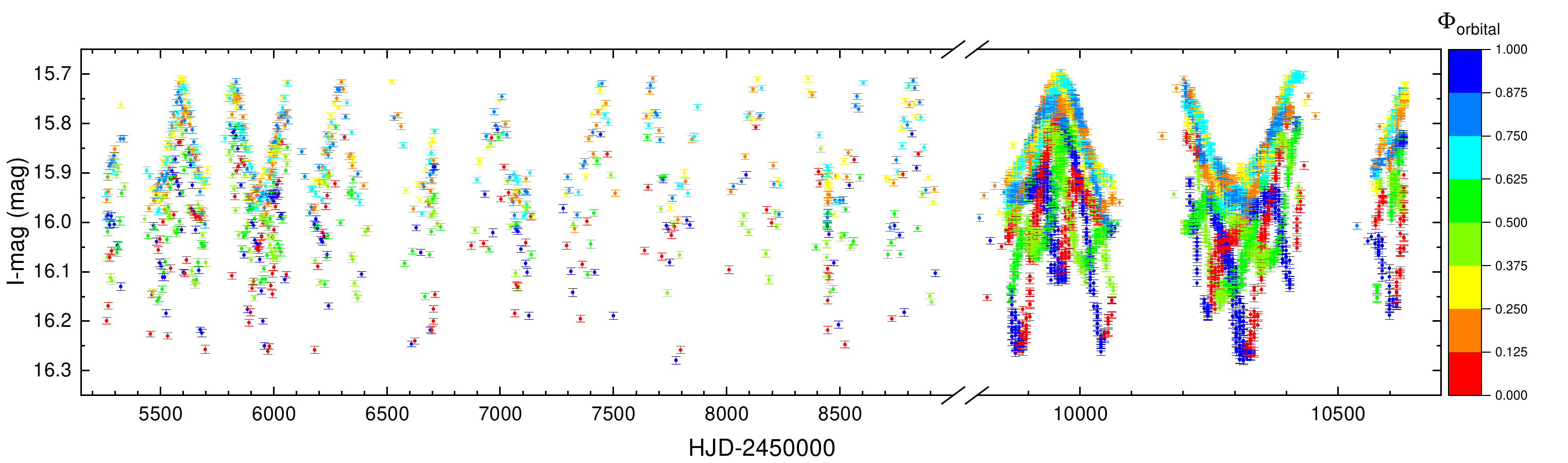}}
\caption{Section of the OGLE $I$-band light curve of \var. Colors indicate different orbital phases.}
\label{fig:Iband}
\end{figure*}

\begin{figure}
\begin{center}
\scalebox{1}[1]{\includegraphics[angle=0,width=8.5cm]{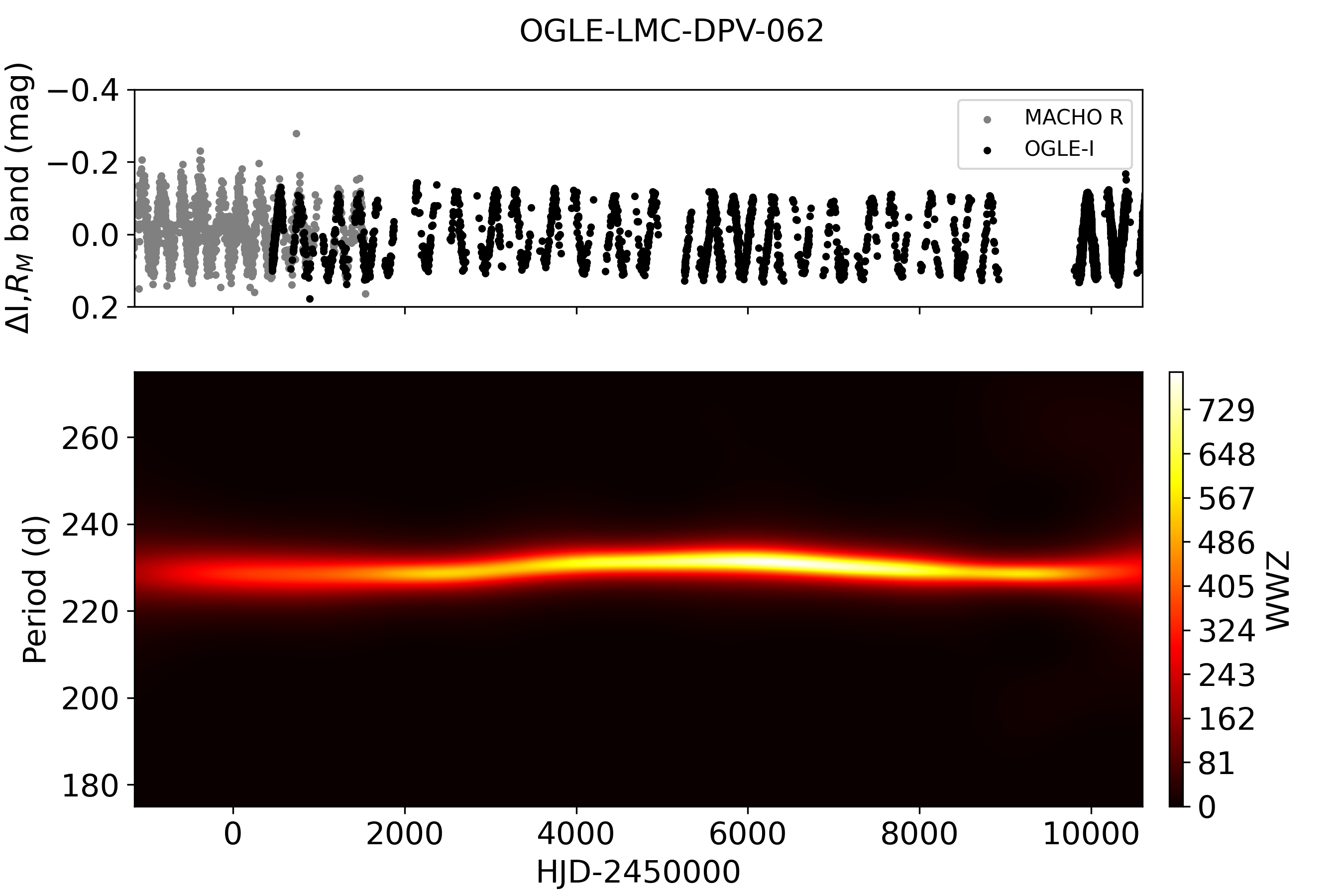}}
\caption{\texttt{WWZ} transform showing a strong signal around 230 days. The orbital period was removed before the analysis.}
\label{fig:wwz}
\end{center}
\end{figure}

\begin{figure}
\begin{center}
\scalebox{1}[1]{\includegraphics[angle=0,width=8.5cm]{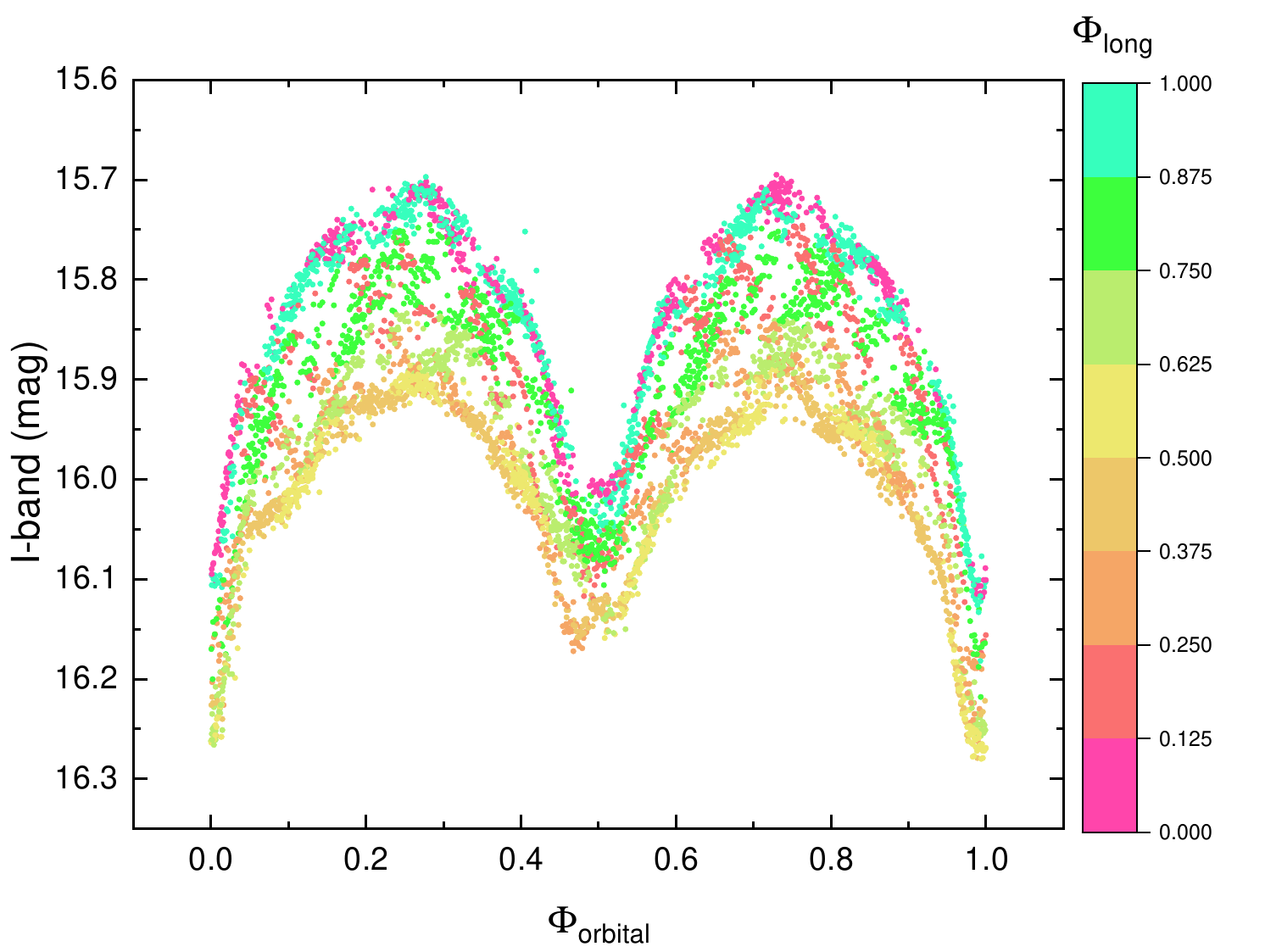}}
\scalebox{1}[1]{\includegraphics[angle=0,width=8.5cm]{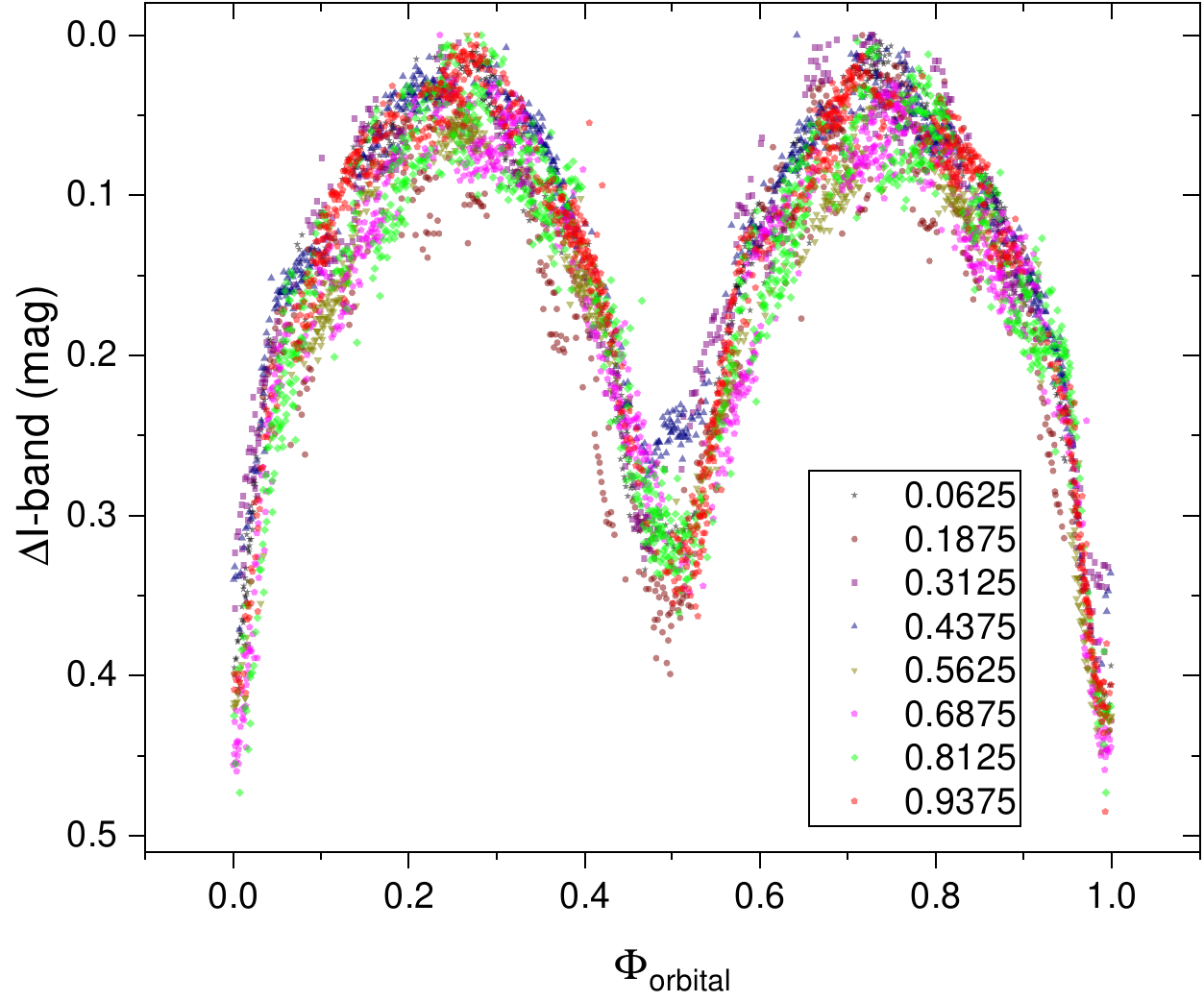}}
\caption{Orbital light curve colored with the long-term cycle phase of the data.  Magnitudes subtracting the minimum magnitude of the respective dataset are shown in the graph below.}
\label{fig:OLCLP}
\end{center}
\end{figure}

\begin{figure}
\begin{center}
\scalebox{1}[1]{\includegraphics[angle=0,width=8.5cm]{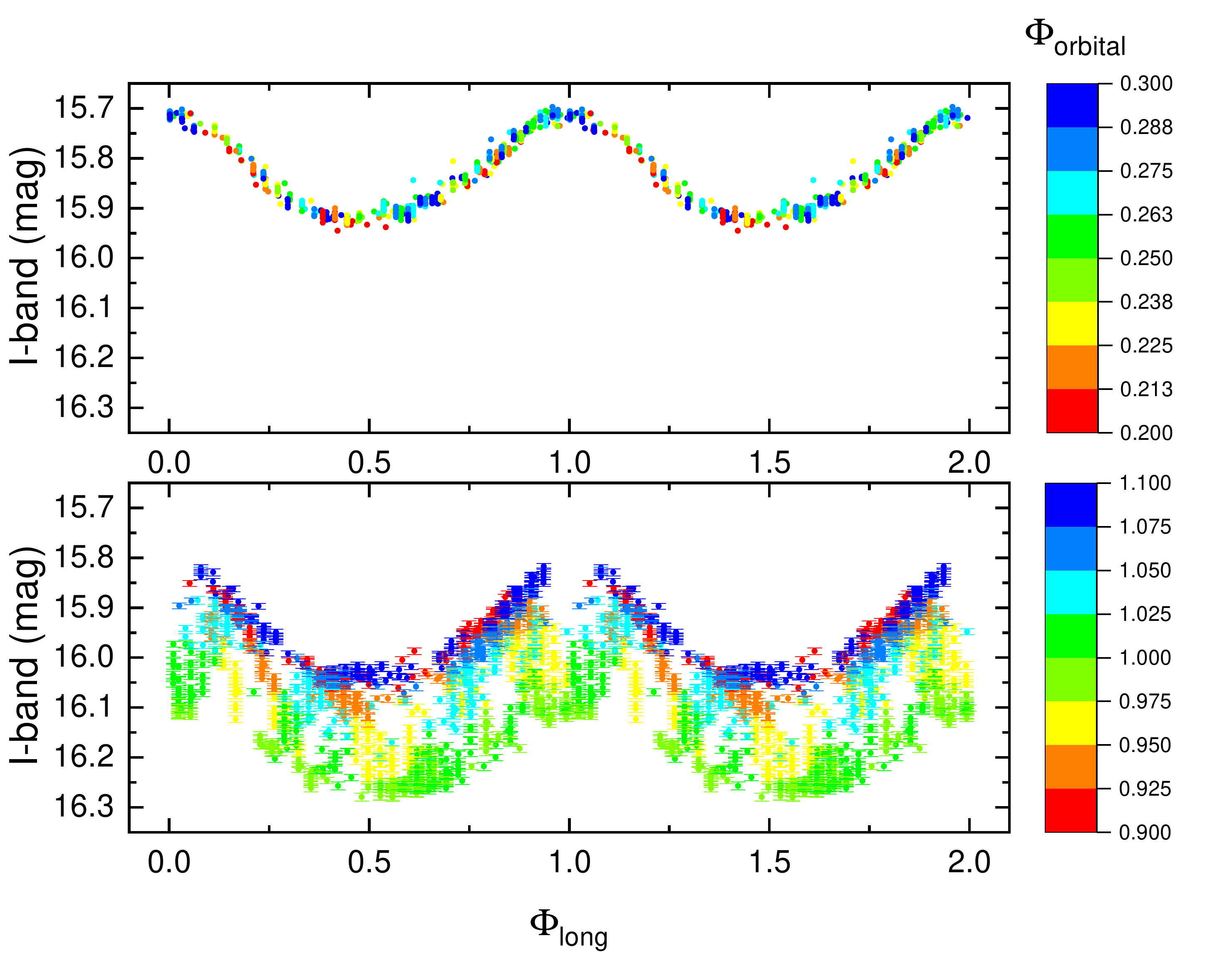}}
\caption{$I$-band magnitudes taken at orbital phases [0.2-0.3; up] and [0.9-1.1; down] phased with the long-term cycle phase. }
\label{fig:LongCycle}
\end{center}
\end{figure}

\begin{figure}
\scalebox{1}[1]{\includegraphics[angle=0,width=8.5cm]{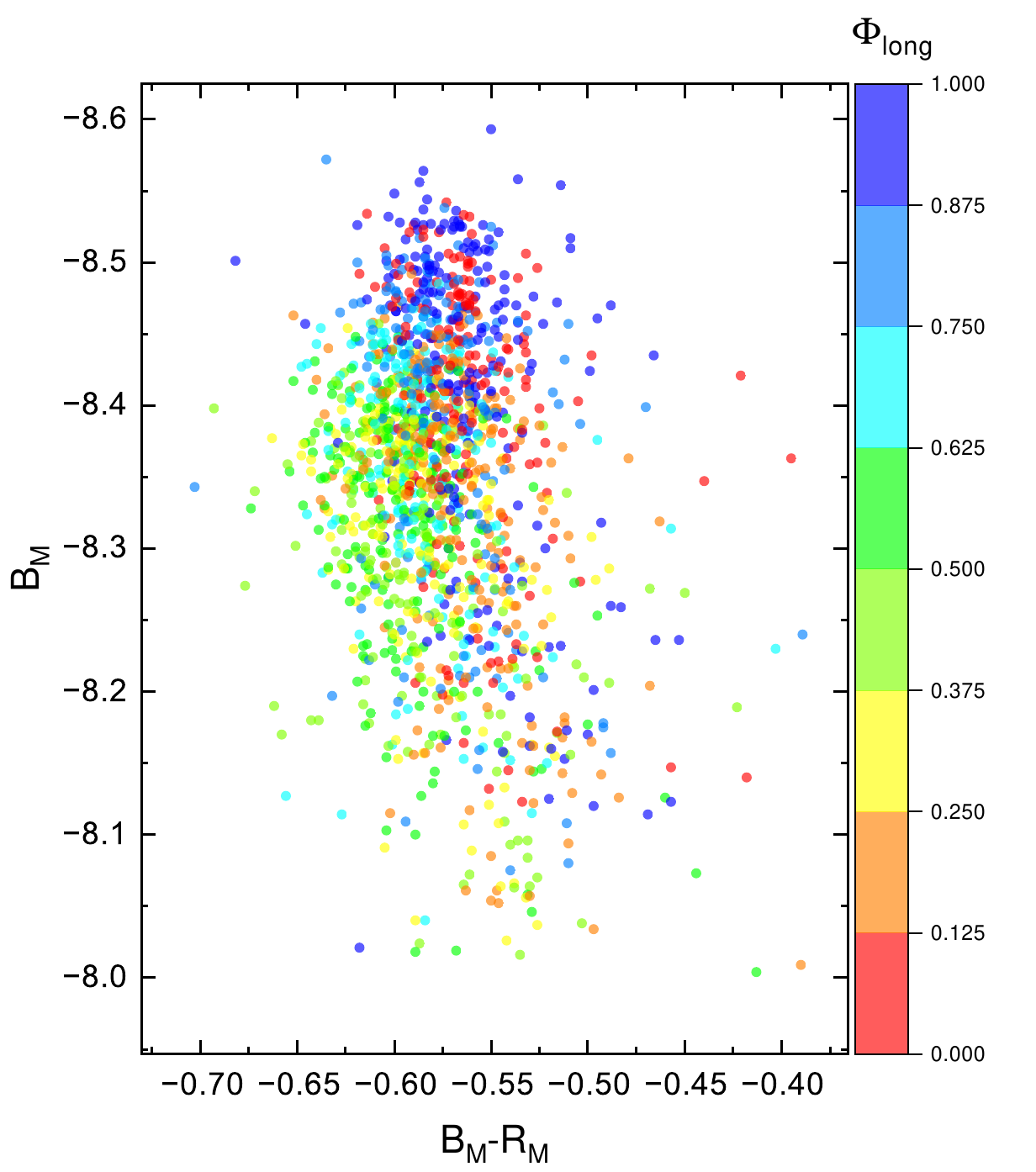}}
\caption{MACHO $B_M$-$R_M$  color versus orbital and long-term cycle phases.}
\label{fig:loop}
\end{figure}

\begin{table}
\centering
\caption{Photometric observations.}
\tiny
\label{tab:observinglog}
\begin{tabular}{lrrrrr} 
\hline
\small 
Source & band &  N      & $\rm{HJD'_{start}}$      &  $\rm{HJD'_{end}}$   & mean $\pm$ std (mag) \\
\hline
MACHO& $B_{M}$&1606&-1173.8652 & 1545.0552 &-8.348 $\pm$ 0.107\\
MACHO&$R_{M}$&1557&-1173.8652 & 1545.0552 &-8.072 $\pm$ 0.104\\
OGLE& $I$ &8942 &455.6827 &10629.8589&15.921  $\pm$ 0.122 \\
OGLE& $V$ &495 &467.7385 &10627.8412& 15.968 $\pm$ 0.110 \\
\hline
\vspace{0.05cm}
\end{tabular} \\
Note: Summary of OGLE and MACHO photometric observations  analyzed in this paper. The number of measurements, starting and ending times, and average magnitude and their standard deviation are given. HJD' = HJD-2450000. The uncertainty of a single measurement is in average 0.006 mag and 0.004 mag for OGLE  $I$ and $V$ magnitudes, and 0.018 mag and 0.013 mag for MACHO $R_{M}$ and $B_{M}$  magnitudes. 
\end{table}

\section{Data and methodology}

\subsection{Photometric data}

The photometric time series analyzed in this study consists of 8942 $I$-band data points and 495 $V$-band datapoints obtained by the 
Optical Gravitational Lensing Experiment \citep[OGLE,][]{2015AcA....65....1U}. The $I$-band light curve shows the typical variability of an eclipsing binary, plus  a remarkable long-term variability (Fig. \ref{fig:Iband}). In addition, we study 1606 $B_{M}$ and 1557 $R_{M}$ data points obtained  by the MAssive Compact Halo Objects (MACHO) project\footnote{https://datacommons.anu.edu.au/DataCommons/item/anudc:3255}.  The two MACHO passbands are nonstandard. The blue band covers 437--590 nm with an effective wavelength of about 520 nm, and the red band covers 590--780 nm with an effective wavelength of about 690 nm. A detailed description of the MACHO photometric system is given by \citet{1997ApJ...486..697A}. The whole dataset, summarized in Table \ref{tab:observinglog}, spans a time interval of  32.3 yr.

\subsection{Time series analysis}

We use the Phase Dispersion Minimization  \citep[\texttt{PDM,}][]{1978ApJ...224..953S} and the Generalized Lomb Scargle software \citep[\texttt{GLS,}][]{2009A&A...496..577Z} to detect periodicities in the light curve.  Long-term tendencies were studied with the weighted wavelet Z transform \texttt{WWZ} 
as defined by \citet{1996AJ....112.1709F}. Each $I$ and $R_{M}$ light curve was shifted to a common mean magnitude before applying this transform. The \texttt{WWZ} works in a similar way to the Lomb-Scargle periodogram providing information about the periods of the signal and the time associated with
those periods. It is very suitable for the analysis of non-stationary signals and has advantages for the analysis of time-frequency
local characteristics.  In order to search for the long-term cycle length, we removed the contribution of the orbital cycle from the observed light curve by fitting a Fourier series to the photometry. This fit was performed using the orbital phase, calculated from the system's ephemeris, and the observed magnitude as a function of phase. 
The Fourier function included 16 harmonics, which allowed for an accurate reproduction of the orbital light curve morphology. The fit was carried out using the \textit{curve-fit} function from \textit{scipy.optimize}, optimizing the coefficients of the Fourier series. Once the optimal parameters were obtained, the fitted model was calculated, and the observed magnitudes were subtracted, yielding the residuals that entered into the  \texttt{WWZ} algorithm.

\subsection{The light curve model}

The inverse problem was  solved using a sophisticated simplex algorithm \citep{DT91}, allowing us to fit the observed light curve by optimizing the parameters of the star$-$orbit$-$disk configuration. The foundations of this method and the procedure for generating synthetic light curves are extensively described in the literature \citep{1992Ap&SS.196..267D, 1996Ap&SS.240..317D}, with further enhancements reported by \citet{2008AJ....136..767D}. This modeling approach has been successfully applied in the analysis of numerous interacting binaries \citep[e.g.,][]{2013MNRAS.432..799M, 2018MNRAS.476.3039R, 2020A&A...642A.211M}.

The total flux from the system is modeled as the sum of the stellar contributions plus the emission from an optically thick accretion disk surrounding the more luminous component, with projection effects modulated by the orbital inclination. The disk's emission is calculated using local Planck functions at characteristic temperatures,  neglecting detailed radiative transfer. Nonetheless, the model accounts for reflection effects, limb darkening, and gravity darkening  of the stars.
 
The accretion disk is described by its radius $R_d$, its vertical semi-thicknesses at the center and outer edge ($d_c$ and $d_e$, respectively), and a radially varying temperature profile \citep[e.g.,][]{2021A&A...645A..51B}:

\begin{equation}
T(r) = T_{d} \left(\frac{R_{d}}{r}\right)^{a_{T}}, 
\end{equation}

\noindent where $T_d$ denotes the temperature at the disk's outer rim ($r = R_d$), and $a_T$ is the radial temperature gradient exponent, constrained to $a_T \leq 0.75$. When $a_T = 0.75$, the profile corresponds to a steady-state configuration. This temperature law reflects a hotter inner disk region that gradually cools toward the outer edge.

 The model also includes two localized active regions at the disk rim: the hot spot, placed near the expected impact site of the gas stream from the inner Lagrangian point, and the bright spot, located at a different azimuth. Both regions are assumed to be hotter and vertically more extended than the surrounding disk, in agreement with Doppler tomography and hydrodynamical studies of interacting binaries \citep[e.g.,][]{1996ApJ...459L..99A,2000A&A...353.1009B,2012ApJ...760..134A}. Similar structures have also been inferred in the DPV $\beta$ Lyrae from light-curve modelling, interferometry, and polarimetry \citep{2013MNRAS.432..799M,2012ApJ...750...59L,2018A&A...618A.112M}. While the hot spot is naturally associated with the stream–disk interaction, the bright spot may arise from gas that has crossed the stream–disk shock/discontinuity region (the so-called hot line) and then flows along the outer disk rim with enhanced vertical motion. Hydrodynamical calculations show that this process can produce vertical oscillations and secondary thickened regions at other azimuths, providing a plausible physical origin for the bright spot \citep{2017ARep...61..639K}.

Each spot is characterized by a relative temperature, $A_{hs} \equiv T_{hs}/T_d$ for the hot spot and $A_{bs} \equiv T_{bs}/T_d$ for the bright spot, their angular extents ($\theta_{hs}$ and $\theta_{bs}$), and azimuthal positions with respect to the line of centers in the direction of orbital motion ($\lambda_{hs}$ and $\lambda_{bs}$). The parameter $\theta_{rad}$ denotes the angle between the local disk surface normal and the direction of peak radiation from the hot spot.

 We acknowledge that the model described above is a simplified representation of the disk rather than a comprehensive physical simulation. Nevertheless, as we will demonstrate later, it enables the extraction of certain physical parameters and the description of the system photometric variability in terms of variations in disk parameters.

For gravity darkening, the coefficients are set to $\beta_1 = 0.25$ and $\beta_2 = 0.08$, corresponding to radiative and convective envelopes, respectively, while the albedos are ${\rm A_1 = 1.0}$ and ${\rm A_2 = 0.5}$, in line with von Zeipel's law for radiative flux redistribution \citep{1924MNRAS..84..702V}. 
 Limb darkening was modeled using the four-coefficient non-linear law of \citet{2000A&A...363.1081C}, with the passband-specific coefficients recalculated at each iteration by bilinear interpolation in $T_{eff}$ and log\,g from tabulated values for the relevant photometric bands; the same prescription was also applied to the disk \citep{2010MNRAS.409..329D}.

In addition, we calculated the  relative mass transfer rate $\dot{M}$ using the approximation given by \citet{2021A&A...653A..89M}

\begin{eqnarray}
 \frac{\dot{M}_{2,f}}{\dot{M}_{2,i}}= \frac{  {{R}^2_{disk,f}} [A_{hs,f}T_{disk,f}]^{4} d_{e,f}  \theta_{hs,f} }{{{R}^2_{disk,i}} [A_{hs,i}T_{disk,i}]^{4} d_{e,i} \theta_{hs,i}}.
\end{eqnarray} 

\noindent This formula assumes a hot spot along the disk border whose luminosity is due to the release of gravitational energy when the gass stream coming from the inner Lagrangian point impacts its surface. This approximation enables us to calculate the relative mass transfer rates at various times, denoted by epochs $i$ and $f$, within a specific system. 

\subsection{The evolutionary code}

For the calculation of the evolution of the possible DPV062 progenitor binary systems, the \texttt{MESA} 
(Modules for Experiments in Stellar Astrophysics) code (Paxton et al 2011, 2013, 2015, 2018) was used, in revision 10398.
This code calculates the detailed evolution of both stars. The stellar wind mass loss rate of each star is calculated according to  Vink et al. (2001) for stars with surface abundance of hydrogen above 0.4 and Nugis \& Lamers (2000) for the hydrogen abundance below 0.4. The matter lost due to the stellar wind has the specific orbital angular momentum of its star. Mass transfer rate is calculated according to  Ritter (1988). The composition of accreted material is identical to the donor's current surface composition. The metallicity is set to the LMC value of 0.006 (Eggenberger et al. 2021).

\section{Results}

\subsection{Period analysis}

Using the $I$ band photometry, and  using the \texttt{PDM} algorithm, we find  the following ephemeris for the occultation of the cooler star:

\small
\begin{eqnarray}
\rm HJD&=&\rm 5255\fd4980   \pm 0\fd0138 + (\rm 6\fd904858 \pm 0\fd000015) \, E.
\end{eqnarray}
\normalsize

This finding was confirmed applying the  \texttt{GLS} and it is consistent with the previous value given by \citet{2010AcA....60..179P}. The analysis of the minima 
of the light curve confirmed that,  for all practical purposes, the orbital period is constant.

We find that the \texttt{WWZ} transform for the $I$-band and $R_M$ time series, suggests a rather stable  long-term cycle length of  around 230 days (Fig. \ref{fig:wwz}).   
We determined the following ephemeris for the maximum of the long cycle:

\small
\begin{eqnarray}
\rm HJD&=&\rm 309\fd9 \pm  2\fd0 + (\rm 229\fd7 \pm 0\fd1) \, E.
\end{eqnarray}
\normalsize

 We note that this long period is longer than that reported by \citep{2010AcA....60..179P}, which may be attributable to the longer time series analyzed in the present study along with possible changes in long-term variability time scale. We find that the shape of the orbital light curve changes according to the long-term cycle phase (Fig. \ref{fig:OLCLP}). The overall  system brightness follows the long-term cycle and 
 the secondary minimum shifts around long-cycle phase 0.5. Changes are observed in the overall shape of the orbital light curve at different long-term cycle phases. The long-term cycle is revealed in slices of data taken at quadrature and primary eclipse (Fig. \ref{fig:LongCycle}). While at quadrature the light curve shows a smooth oscillatory pattern, the data at main eclipse show much larger scatter.

\subsection{MACHO colors and extinction}

The MACHO light curve follows the behaviour shown by the OGLE light curves but with higher scatter (Fig.\, \ref{fig:machoLCs}). The photometry reveals that, except for eclipses, the main color trends are observed during the long cycle, not during the orbital cycle, and that these variations are of rather small amplitude (Fig.\, \ref{fig:machocolor}). The data also show that, in general, the system is redder when brighter, i.e. usually near the maximum of the long cycle. In addition, a color loop is observed, the rising branch shows bluer colors compared to the declining branch  (Fig.\, \ref{fig:loop}),  something also reported in the DPV OGLE05155332-6925581 \citep{2008MNRAS.389.1605M}.

For the coordinates of OGLE-LMC-DPV-062 we obtained  $E(V-I)$ = 0.073 $\pm$ 0.100 following extinction maps in the Large Magellanic Cloud  \citep{2021ApJS..252...23S}.
The extinction $A_I$ can generally be estimated as $A_I$ = 1.5 $\times$ $E(V-I)$  \citep{2021ApJ...919...99I}. For our case, this is $A_I$ = 1.5 $\times$ 0.073 = 0.11  $\pm$ 0.15.  Using  $A_V$ = (2.5 $\pm$ 0.2) $\times$ $E(V-I)$  \citep{2021ApJ...919...99I}, we get $A_V$ = 0.183 $\pm$ 0.015. We did not found extinction transformation coefficients for MACHO R- or B-bands. Due to the complex nature and uncertainties involved in the MACHO photometry and the atypical bandpass filters  \citep{1999PASP..111.1539A}, we consider MACHO photometry only for the light curve models and eclipse timings.

\begin{figure}
\begin{center}
\scalebox{1}[1]{\includegraphics[angle=0,width=8.5cm]{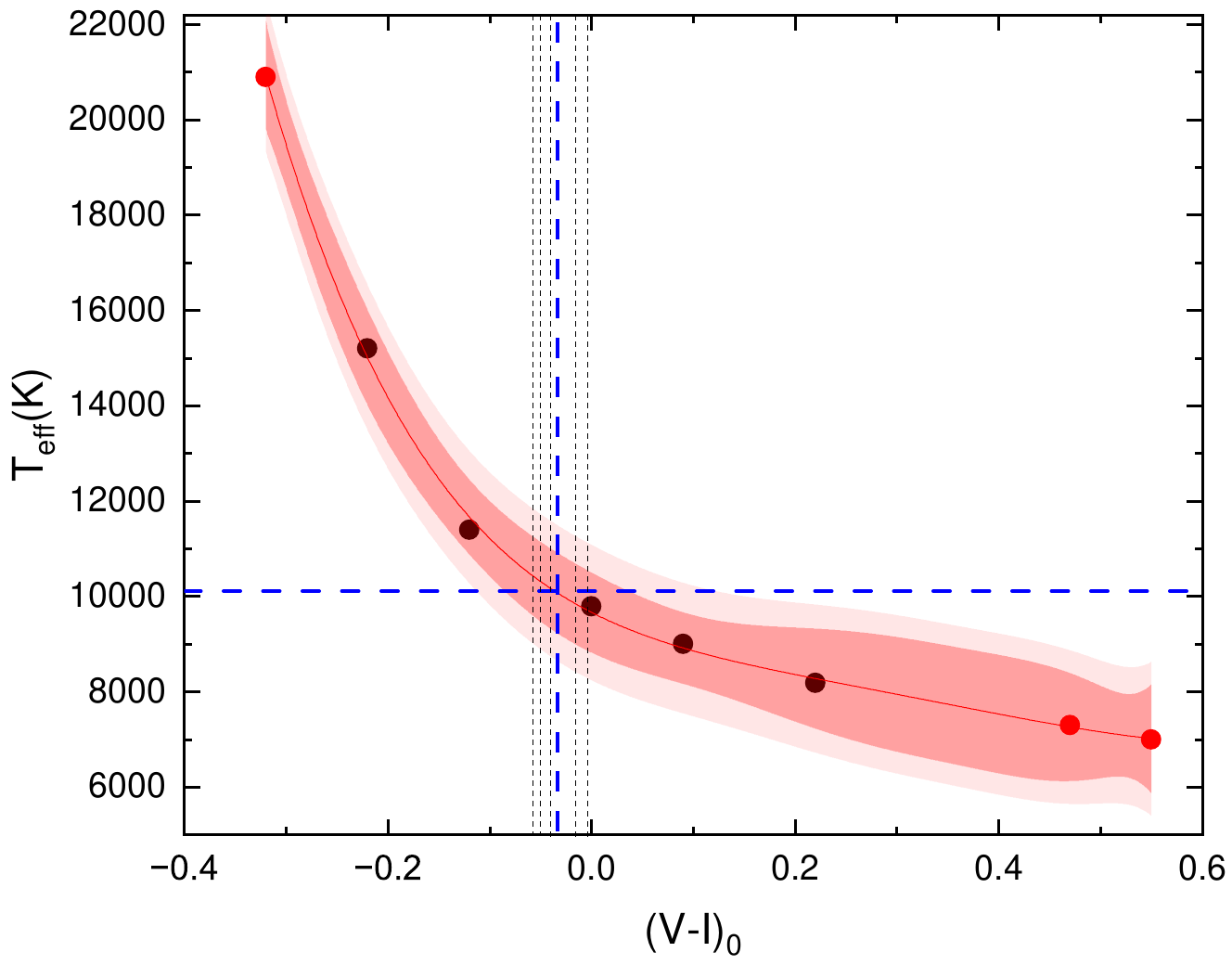}}
\caption{Dereddened colors and effective temperatures for giant (red dots) and dwarfs (black dots). 
The best 3th order polynomial fit along with 95\% confidence and prediction bands are shown by dashed
and dashed-light areas, respectively.  We show the observed colors of \var\, during main eclipse in five long-cycle phases as vertical lines, while the solid dashed line shows the average color and the horizontal line the upper limit for the donor temperature, viz., 10111 $\pm$ 308 K  (see text for details).  }
\label{fig:T2}
\end{center}
\end{figure}

\subsection{The stellar parameters}

The absence of spectroscopic data limits the accuracy of our knowledge of the stellar parameters; however, we can still obtain preliminary estimates, bearing in mind that spectroscopic studies are required to confirm them. Nevertheless, the general trends and patterns derived from the analysis of the long cycle are expected to remain robust against small variations in the adopted stellar parameters.

If the main eclipse is total, then the color observed at this epoch corresponds to the donor. As our system shows partial eclipses, the gainer (and eventually the disk)  also contributes to the flux at minimum.  This means that only an upper limit for the donor effective temperature can be derived from the system color at main eclipse. After a careful inspection of the $I-V$ band light curves during main eclipse (between $\Phi_{\rm o}$ = 0.95 and 1.05) around different long-term phases ($\Phi_{\rm l}$  = 0.0, 0.4, 0.6, 0.7 and 0.8), we find $V-I$ = 0.039 $\pm$ 0.023. Introducing the aforementioned reddening correction we arrived to the un-reddened color $(V-I)_{0}$ = -0.0338 $\pm$ 0.1019.
Using the color-temperature calibration described in \citet{2020A&A...641A..91M} we get the upper limit $T_{2(\rm{upper\, limit})}$ = 10111 $\pm$ 308 K (Fig. \ref{fig:T2}).

\begin{figure}
\begin{center}
\scalebox{1}[1]{\includegraphics[angle=0,width=7cm]{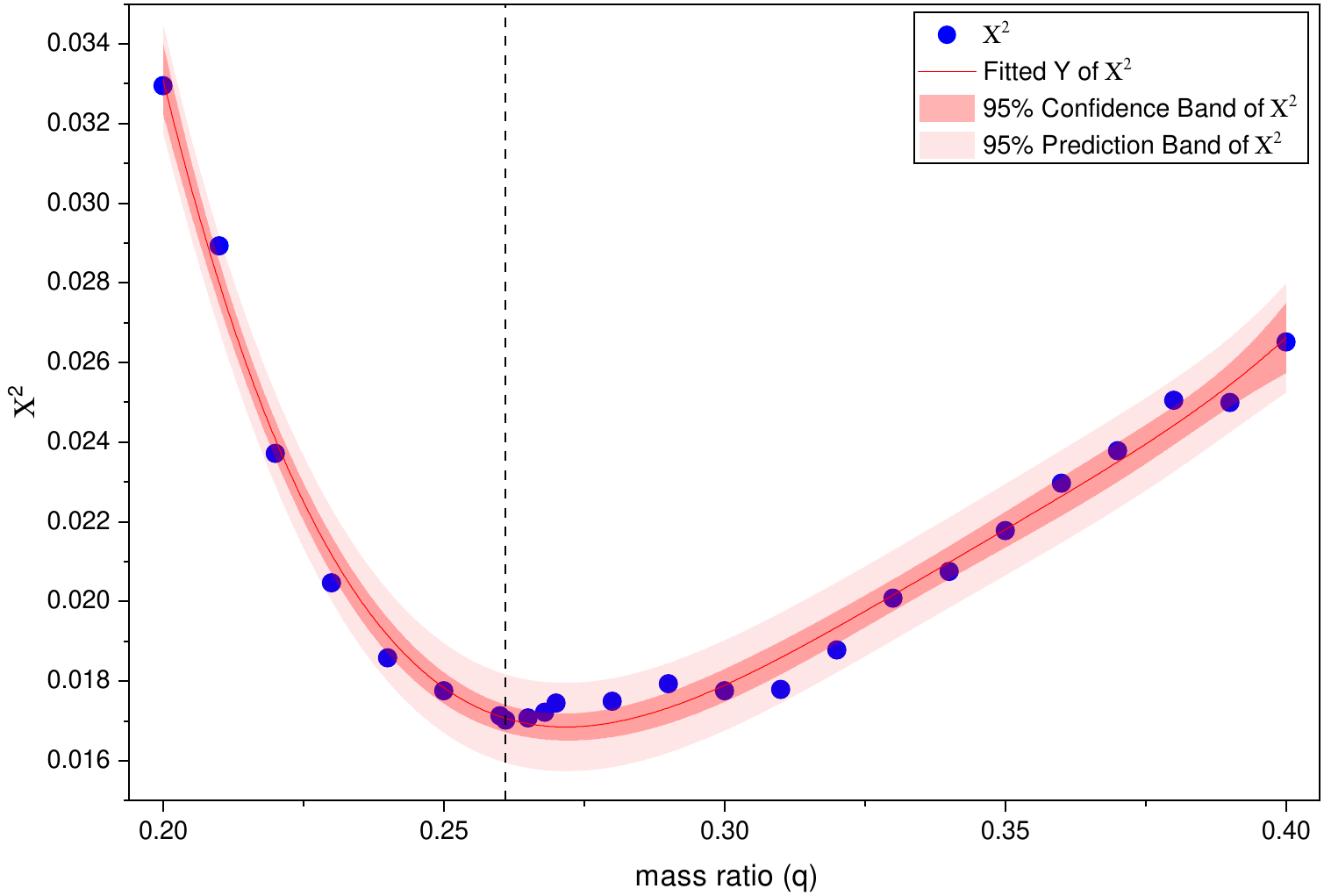}}
\caption{Parameter S = $\Sigma$ (O-C)$^{2}$ for the fits done to the light curve of dataset 1, as a function of mass ratio. The best 4th order polynomial fit is also shown along with a vertical line showing the minimum at $q$ = 0.261. }
\label{fig:qsearch}
\end{center}
\end{figure}

In the analysis of over-contact or semi-detached binary systems based solely on photometric time series and lacking spectroscopic data, the q-search method is widely used to estimate the mass ratio of the system. This technique is particularly reliable for eclipsing binaries  showing total eclipses \citep{2005Ap&SS.296..221T},  and might be considered an approximation for high inclination binaries, although resulting uncertainties may still be underestimated. 
By applying this method, we obtained convergent solutions for a range of mass ratios defined as $q = M_2/M_1$, where subscripts 2 and 1 refer to the cooler and hotter components, respectively. We selected dataset 1 for this analysis, as it best represents the orbital light curve with minimal brightness attenuation due to disk obscuration of the primary star.
The q-search was conducted over the range $q = 0.20-0.40$, with a variable step size typically of 0.01. A mass ratio of $q$ = 0.261 yielded the best agreement with the observed data (Fig. \ref{fig:qsearch}). This mass ratio is consistent with previous results for Double Periodic Variables (DPVs), where an average value of $q = 0.23 \pm 0.05$ (standard deviation) has been reported \citep{Mennickent2016}. We notice that the minimum of the 4th order polynomial fit is found at 0.271 $\pm$ 0.005, obtained with the bootstrapping technique and 1000 iterations.  Considering the asymmetrical shape of the fit we will consider the mass ratio as $q$ = 0.26 $\pm$ 0.04 as the most reasonable value. From our analysis we also determined that the fits are sensitive to the adopted inclination, and that the best fit is found at inclination angle of about 78 degree.

We calculated stellar and orbital parameters by keeping fixed the mass ratio and starting with a donor temperature of 10kK and then searching for the best fit of the light curve of dataset 1, varying the gainer temperature. The results were compared with the evolutionary track obtained with the \texttt{MESA} code (section 4), finding some miss-matches that were solved lowering the donor temperature and the overall system mass.  The adopted parameters are listed in Table \ref{tab:system} and their errors are based on the estimates provided in Appendix A.

\subsection{Light curve models}

In this section, we present our modeling of the light curve focused on OGLE-IV photometry, as the uncertainties associated with MACHO data are approximately an order of magnitude higher. To investigate photometric variability, the full dataset was partitioned into 20 subsamples, each representing consecutive phases of the long-cycle  (Table \ref{tab:strings}).

\begin{table}
\caption{Data  sets used in this paper. }
\label{tab:strings}
\centering
\begin{tabular}{l r r c  r r }
\hline
\hline
set&N & I$_{0.25}$ &$\Delta$I$_{0.25}$ (mag)  &$\Phi_{\rm{L}}$ & $\dot{M}$ \\
\hline
01 &295&15.711	&0.000	&0.025&	3.41    \\
02 &237&15.728	&0.017     &0.075&	3.34    \\ 
03 &255&15.754	&0.043	&0.125&	1.61   \\
04 &244&15.785	&0.074	&0.175&	1.18    \\
05 &235&15.829	&0.118     &0.225&    2.067    \\
06 &250&15.864	&0.154	&0.275&   7.37    \\
07 &219&15.890	&0.179	&0.325&	5.15    \\
08 &314&15.903	&0.192	&0.375&10.20    \\
09 &321&15.914	&0.204	 &0.425&	6.71     \\
10 &247&15.913	&0.202	&0.475&	12.22   \\
11 &201&15.904	&0.193	 &0.525&	16.25  \\
12 &300&15.908	&0.198	&0.575&	8.61   \\ 
13 &366&15.897	&0.186	&0.625&	11.01   \\
14 &338 &15.875	&0.165	&0.675&	3.22 \\
15 &354 &15.841	&0.130	&0.725&	3.30 \\
16 &382 &15.820	&0.109	&0.775&	2.17 \\
17 &470&15.788	&0.077	&0.825&	1.04\\
18 &459 &15.747	&0.036	&0.875&	1.00\\
19 &442&15.725	&0.015	&0.925&   3.14\\
20 &300&15.712	&0.001	&0.975&   5.98\\
\hline      
\end{tabular}
\tablefoot{The magnitude at orbital phase 0.25, referred to data set 1, is shown according to the light-curve model. The mass-transfer rates, normalized to the value of data set 18 and defined in Sect. 3.2, are also given, with typical uncertainties of about 25\%. The mean phases of the long-term cycle ($\Phi_{\rm L}$) and the number of data points (N) are provided as well.}
\end{table}

\begin{table}
\centering
\caption{Summary of stellar and orbital parameters and their formal errors.  }
\label{tab:system}
\begin{tabular}{lclc} 
\hline
\hline
Parameter & Value & Parameter &Value \\
\hline
 $M_1$  (M$_{\odot}$) & 10.36  $\pm$ 1.14 & log \ $g_1$    & 4.03  $\pm$  0.17 \\
 $M_2$  (M$_{\odot}$) & 2.73  $\pm$  0.30 & log \ $g_2$    & 2.90  $\pm$   0.07 \\
 $R_1$ (R$_{\odot}$)  & 5.14  $\pm$ 0.98 &$P_{\rm{o}}$  (d) &6.904858(10)  \\
 $R_2$ (R$_{\odot}$)  & 9.71  $\pm$   0.56 &   $a_{\rm{orb}}$ (R$_{\odot}$)    & 35.9 $\pm$  1.4    \\
 $T_1$  (K) &    24400 $\pm$  2400     & $i$ ($^{\rm{o}}$) &      77.7  $\pm$  0.1\\
  $T_2$  (K)  &  7100 $\pm$  700     & & \\     
  \hline
\end{tabular}
\tablefoot{Indexes 1 and 2 refer to hot and cool stellar components. The error in the orbital period is given in parentheses for the three last digits.}
\end{table}
\normalsize

\begin{table*}
\centering
\caption{Light curve fit parameters for the data segments at the $I$-band along with formal errors, mean and standard deviation.  Mean results are also given for the analysis at the $B_{M}$ and $R_{M}$ bands. }
\label{tab:results}
\begin{tabular}{crrrrrrrrrrrr}
\hline 
\hline
set&   $T_{disk}$&   $A_{hs}$&  $\theta_{hs}$ & $\lambda_{hs}$ & $\theta_{rad}$ &  $A_{bs}$&  $\theta_{bs}$ & $\lambda_{bs}$ &  $a_T$   & $R_{disk}$ & $d_e$& $d_c$\\
&  $\pm$300 & $\pm$0.05 & $\pm$0.4 & $\pm$ 9.0 & $\pm$7.0 & $\pm$0.10 & $\pm$3.0 &$\pm$ 14.0  & $\pm$0.05 & $\pm$1.0 & $\pm$1.2 & $\pm$0.5\\
  &(K)&&(\dg)&(\dg)&(\dg)&&(\dg)&(\dg)&&(\rsun)&(\rsun)&(\rsun) \\
\hline
1	&5704&	2.19&	25.2&	314.5&	-17.8	&1.78&	52.7&	62.3&	0.64&	14.73&	3.58&	1.01\\
2	&5397&	2.29&	27.0&	316.3&	-24.5	&1.89&	57.7&	53.2&	0.73&	14.31&	3.54&	1.02\\
3	&5098&	2.06&	24.6&	319.7&	-14.6	&1.69&	47.3&	52.0&	0.54&	16.88&	3.98&	0.74\\
4	&5684&	1.99&	22.2&	321.1&	-13.9	&1.57&	43.5&	51.8&	0.49&	13.71&	5.37&	0.68\\
5	&5195&	2.10&	23.3&	323.1&	-13.2	&1.62&	44.1&	56.0&	0.53&	14.47&	4.61&	1.18\\
6	&5010&	2.18&	25.1&	314.6&	-11.2	&1.83&	42.4&	45.0&	0.36&	17.77&	2.48&	2.60\\
7	&5144&	2.23&	23.4&	325.4&	-16.2	&1.89&	34.3&	52.6&	0.33&	14.45&	3.34&	2.40\\
8	&5357&	2.27&	25.4&	330.0&	-25.7	&1.88&	45.9&	51.7&	0.59&	13.79&	3.11&	3.80\\
9	&4896&	2.26&	25.6&	328.0&	-18.7	&1.84&	57.1&	47.8&	0.69&	14.19&	2.95&	3.42\\
10	&5392&	2.29&	26.1&	328.6&	-13.4	&1.71&	55.3&	56.8&	0.73&	13.53&	2.99&	4.38\\
11	&5485&	2.29&	26.7&	330.5&	-15.0	&1.70&	57.5&	66.3&	0.75&	13.94&	3.09&	4.95\\
12	&5038&	2.29&	25.2&	332.2&	-20.5	&1.74&	57.8&	61.5&	0.71&	14.10&	3.36&	3.82\\
13	&5524&	2.28&	25.6&	329.5&	-24.6	&1.88&	46.4&	58.8&	0.55&	13.31&	3.24&	3.81\\
14	&4788&	2.25&	23.2&	329.0&	-20.5	&1.88&	42.5&	57.6&	0.26&	15.08&	3.37&	1.79\\
15	&5017&	2.24&	24.9&	322.3&	-14.8	&1.78&	56.1&	58.9&	0.42&	14.19&	3.26&	1.62\\
16	&5058&	2.15&	24.4&	324.7&	-16.9	&1.75&	49.0&	55.0&	0.45&	14.84&	3.74&	1.13\\
17	&5143&	2.16&	25.0&	322.5&	-17.2	&1.75&	50.5&	60.8&	0.51&	13.97&	4.14&	0.55\\
18	&5411&	2.12&	22.4&	320.8&	-15.0	&1.63&	56.4&	60.5&	0.61&	14.44&	4.55&	0.49\\
19	&5227&	2.29&	26.9&	318.1&	-17.0	&1.84&	56.6&	61.5&	0.75&	15.04&	3.54&	0.99\\
20	&5830&	2.29&	26.6&	316.0&	-17.5	&1.84&	56.9&	61.6&	0.70&	14.05&	3.34&	1.41\\
\hline 
Mean ($I$)    &5270 &2.21&24.9   &323.3&-17.4&1.77&50.5&56.6&0.57&14.54&3.58&2.09   \\
std                      &282  &0.09 &1.4    &5.7     &4.0    &0.10&6.9  &5.4 &0.15&1.07  &0.67&1.44\\
\hline
Mean ($R_{M}$)& 4551 &2.25   &26.3 &327.6 &-13.0&1.86  &52.9 &49.6 &0.64&13.52 &3.80&3.87\\
std                    & 462   &0.07   &0.7   &8.4     &9.2    &0.07  &6.3   &5.4  &0.17&1.34   &1.31&0.75\\
\hline
Mean ($B_{M}$))&5225 &2.17&25.7&317.8&-19.6&1.85&51.6&64.6&0.61&14.57&4.15&2.76\\
std                  &241   &0.09&1.2  &5.9&4.8&0.05&6.5&5.2&0.13&0.69&0.65&0.84\\
\hline
\end{tabular}
\end{table*}

We calculated the mass transfer rate according to Eq.\,2. To standardize the values of $\dot{M}$ we obtained, we referenced the minimum value, which was recorded at  data set 18. These values are displayed in Table \ref{tab:strings}.  The  normalized mass transfer rates  calculated in this way have a mean value of 5.45$\times$ with a standard deviation of 4.29$\times$ , and maximum and minimum values of 16.25$\times$  and 1$\times$ . An uncertainty  for $\dot{M}$ of the order of 25\% was estimated considering the scatter of the fit to the long-term tendency discussed in Section 3.4.  By contrast, the uncertainty obtained through error propagation may reach 40\%. The variability of $\dot{M}$ is treated qualitatively in this paper and is not constrained by these limits.  

The stellar and orbital parameters derived in the above section were held fixed for all datasets, allowing us to focus the analysis on the variability of the accretion disk properties. The full results from the light curve fitting are presented in Table~\ref{tab:results}. Parameter uncertainties were estimated from the 95\% confidence bands of the fits of the parameters during the long-term cycle, discussed later in this section,  they are consistent with uncertainties estimated from numerical trials of the parameters around the best-model values. 

Orbital light curve fits for individual datasets are shown in Fig.\ref{fig:LC1} and Fig.\ref{fig:LC2} illustrating measurement uncertainties, typical orbital features, long-term variability, fit quality, and the corresponding visual representations of the system. Discrepancies beyond the formal errors of individual data points may reflect underestimated uncertainties or model limitations. For example, the model does not consider possible additional light sources. Furthermore, since some light curves span multiple orbital periods, long-term variations may also affect the fits. To mitigate this, we applied a low-degree sliding polynomial to each light curve segment and determined magnitudes at orbital phases 0.25 using the corresponding phase values from the polynomial fits (see Table~\ref{tab:strings}). 

Our analysis reveals that the disk radius varies between 13.31~\rsun\ and 17.77~\rsun, while the disk temperature ranges from 4788~K to 5830~K. The outer vertical thickness of the disk spans 2.48$-$5.37~\rsun, and the inner vertical thickness varies between 0.49 and 4.95~\rsun. The temperature of the hot spot is, on average, 2.21~$\pm$~0.09 times higher than the surrounding disk, while the bright spot is 1.77~$\pm$~0.10 times hotter than its surroundings.

The hot and bright spots are located at 323\fdg3~$\pm$~5\fdg7 and 56\fdg6~$\pm$~5\fdg4, respectively, measured from the line connecting the stellar centers (from the donor to the gainer), in the direction of orbital motion. Their angular extents along the outer edge of the disk are 24\fdg9~$\pm$~1\fdg4 for the hot spot and 50\fdg5~$\pm$~6\fdg9 for the bright spot.

The MACHO $B$ and $R$ light curves show similar tendencies, but with much larger noise. We show their average values for their fits in Table~\ref{tab:results}.

\begin{figure}
\scalebox{1}[1]{\includegraphics[angle=0,width=8cm]{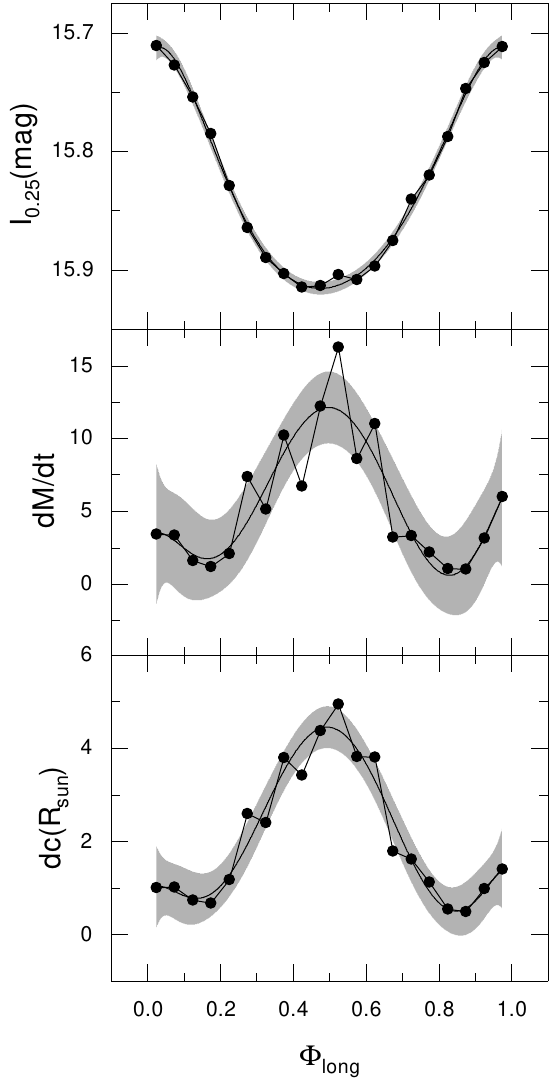}}
\caption{Magnitude, normalized mass transfer rate and disk thickness at the central edge, as a function of the long-term cycle phase. Sixth-order polynomial fits along with the 95\% confidence bands are shown. The parameters of these fits are given in Table \ref{tab:results}. }
\label{fig:LCresults}
\end{figure}

We find that the mass transfer rate is maximum at the long cycle phase 0.5, coinciding with the minimum of the 
magnitude at orbital phase 0.25 and roughly with the minimum of the  disk radius revealed by their fit function (Fig.\,\ref{fig:LCresults}). Most notably, the minimum brightness is found when the inner vertical disk extension is maximum i.e. at larger gainer occultation. This finding shows that in this system, the long cycle is mostly driven by cyclic occultations of the gainer by a disk of variable thickness. 

\begin{figure}
\scalebox{1}[1]{\includegraphics[angle=0,width=8.5cm]{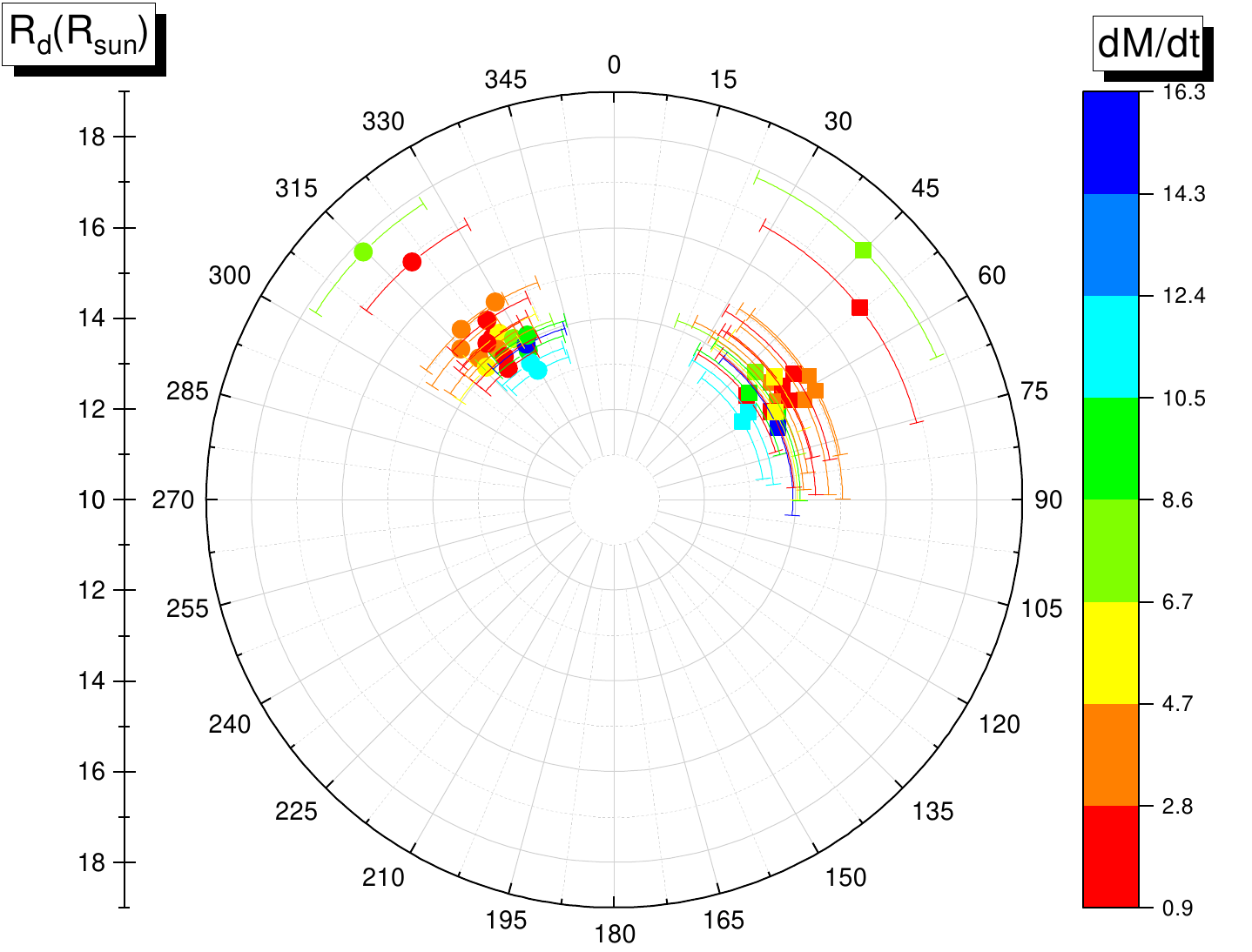}}
\caption{Radial position and angular extension of hot spot (solid circles) and bright spot (squares) at different long-cycle phases. Colors indicate the normalized mass transfer rate.}
\label{fig:polar}
\end{figure}

In addition, the angular positions of the hot and bright spots are constrained to well defined regions during the cycle (Fig.\,\ref{fig:polar}).  The  average radius and temperature of the disk are consistent considering the fits for the $I$, $R_M$ and $B_M$ bandpasses. 

The correlation matrix confirms the previous results and indicates that no parameter degeneracies are present (Fig.,\ref{fig:SigMatrix}). We considered as statistically significant those correlations with p-value equal or under 0.05, i.e. with a probability equal or lower than 0.05 that the correlation (or anticorrelation) is non-existent. The analysis reveals that $\theta_{bs}$ and $\lambda_{hs}$ correlates with a$_{T}$ and system brightness, respectively. The parameters $A_{hs}$ and $A_{bs}$ are anticorrelated with $d_e$.
The matrix also confirms the strong correlation between the disk's inner edge thickness and the system brightness. 

\begin{figure}
\begin{center}
\scalebox{1}[1]{\includegraphics[angle=0,width=8.5cm]{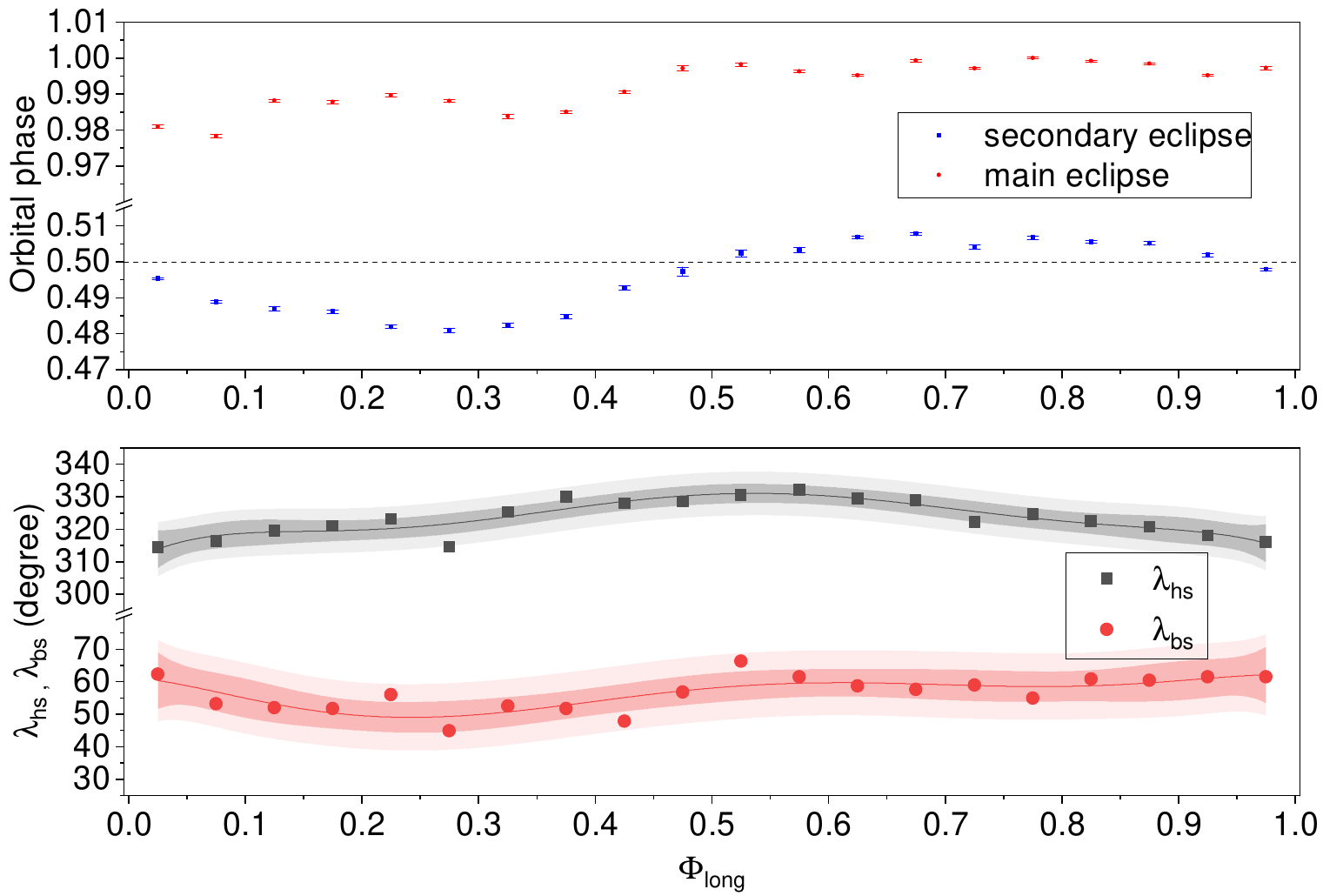}}
\caption{Up: Minima of gaussian fits to the main and secondary eclipses for datasets of Table \ref{tab:strings}. Down: Position of the hot and bright spots during the long cycle and the best 3th-order polynomial fits, along with their 95\% prediction and confidence bands. }
\label{fig:eclipseminima}
\end{center}
\end{figure}

Interestingly, the positions of the main and secondary eclipses, measured with gaussian fits, follow systematic tendencies with the long-term cycle. These changes are reproduced by the models and could indicate changes in the photo-center as suggested by the behaviour of the hot and bright spots during the long cycle (Fig. \ref{fig:eclipseminima}).

\section{On the evolutionary stage}

\begin{figure}
\scalebox{1}[1]{\includegraphics[angle=0,width=8cm]{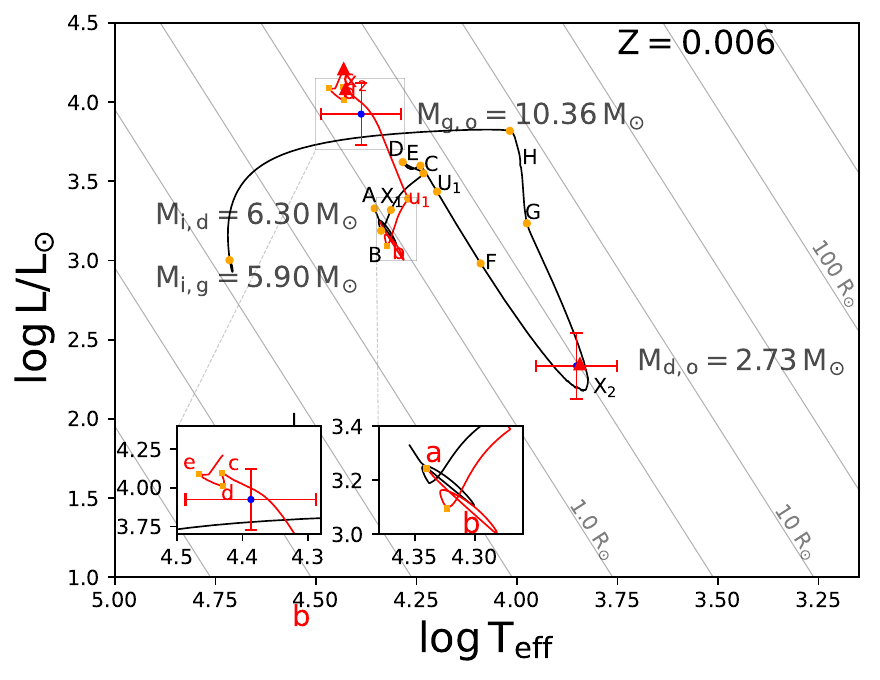}}
\scalebox{1}[1]{\includegraphics[angle=0,width=8cm]{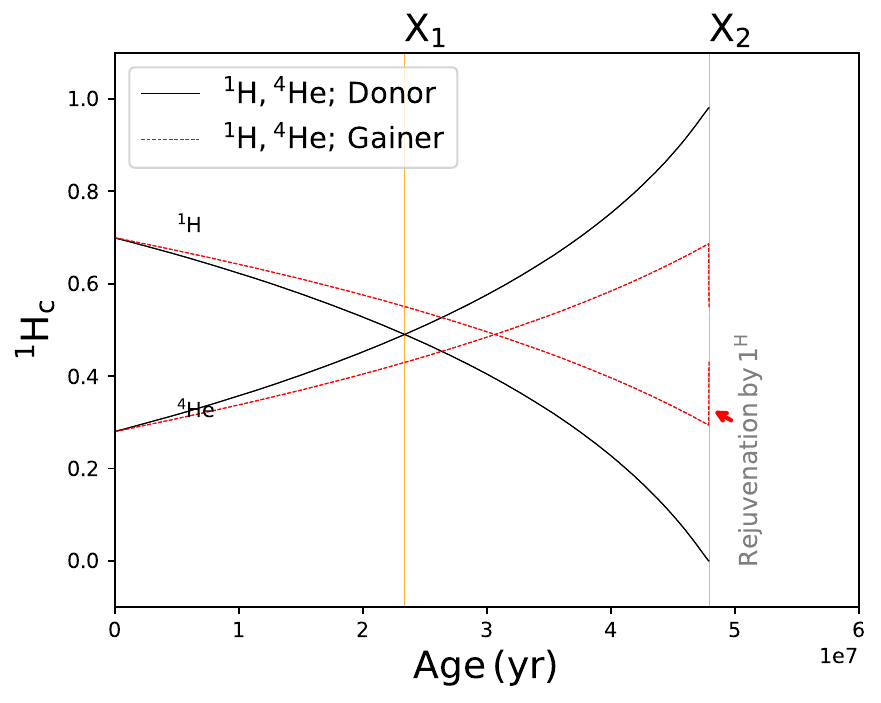}}
\caption{Up: Evolutionary tracks  for the best quasi-conservative model.  Blue dots with error bars are the observed values and triangles 
the model's values. Labels correspond to the evolutive stages given in Table \ref{tab:EVOL}. Down: The central abundance of hydrogen and helium for the gainer and donor. \textrm{X$_{1}$} corresponds to the inversion of the ${}^{1}$H/${}^{4}$He ratio and \textrm{X$_{2}$} to the current stage.}
\label{fig:LT}
\end{figure}

\begin{table*}[h!]
\caption{Evolutionary stages of DPV062}
\normalsize
\begin{center}
\resizebox{0.95\textwidth}{4.2cm}{
\begin{tabular}{llrrrrcrl}
\hline
\noalign
{\smallskip}
\textrm{}       & \textrm{Stage}    & \textrm{Age (Myr)} & \textrm{M (M$_{\odot}$)} & \textrm{R (R$_{\odot}$)}  & \textrm{P$_{\rm o}$ (d)}  & \textrm{$\log{T}$ (K)} & \textrm{$\mathrm{\dot{M}}$\,(M$_{\odot}$ yr$^{-1}$)}   & \textrm{Ev. process}             \\
\hline
\noalign{\smallskip}
\textrm{Donor}  & \textrm{A}        & \textrm{ 0.0000 }  & \textrm{6.3000 }         & \textrm{ 3.0000 }  & \textrm{ 2.5000 }  & \textrm{4.3543} & \textrm{-1.0302E-10 } & \textrm{Zero Age Main Sequence (ZAMS)}\\
\textrm{}       & \textrm{B}        & \textrm{ 0.1548 }  & \textrm{6.3000 }         & \textrm{ 2.7619 }  & \textrm{ 2.5000 }  & \textrm{4.3378} & \textrm{-1.7446E-10 } & \textrm{Terminal Age Main Sequences (TAMS)}\\
\textrm{}       & \textrm{X$_{1}$}  & \textrm{23.3766 }  & \textrm{6.2946 }         & \textrm{ 3.5890 }  & \textrm{ 2.5036 }  & \textrm{4.3136} & \textrm{-3.1368E-10 } & \textrm{Inversion of the ${}^{1}$H/${}^{4}$He ratio}\\
\textrm{}       & \textrm{C}        & \textrm{47.0149 }  & \textrm{6.2825 }         & \textrm{ 6.8059 }  & \textrm{ 2.5113 }  & \textrm{4.2324} & \textrm{-8.5359E-10 } & \textrm{Depletion of central hydrogen ${}^{1}$H}\\ 
\textrm{}       & \textrm{D}        & \textrm{47.8593 }  & \textrm{6.2817 }         & \textrm{ 5.8168 }  & \textrm{ 2.5118 }  & \textrm{4.2841} & \textrm{-1.3862E-09 } & \textrm{Size increase beyond the RL due to depletion of ${}^{1}$H}\\
\textrm{}       & \textrm{E}        & \textrm{47.8839 }  & \textrm{6.2794 }         & \textrm{ 6.9436 }  & \textrm{ 2.5117 }  & \textrm{4.2401} & \textrm{-1.5147E-06 } & \textrm{Initiation of mass transfer}\\
\textrm{}       & \textrm{U$_{1}$}  & \textrm{47.8906 }  & \textrm{6.0816 }         & \textrm{ 6.9598 }  & \textrm{ 2.5059 }  & \textrm{4.1984} & \textrm{-5.3892E-05 } & \textrm{Mass inversion}\\
\textrm{}       & \textrm{F}        & \textrm{47.8993 }  & \textrm{5.3722 }         & \textrm{ 6.8054 }  & \textrm{ 2.6156 }  & \textrm{4.0903} & \textrm{-1.0345E-05 } & \textrm{Minimum Roche lobe value}\\
\textrm{}       & \textrm{X$_{2}$}  & \textrm{47.9236 }  & \textrm{2.8313 }         & \textrm{10.2132 }  & \textrm{ 6.9052 }  & \textrm{3.8434} & \textrm{-9.6518E-05 } & \textrm{Current stage}\\
\textrm{}       & \textrm{G}        & \textrm{47.9413 }  & \textrm{2.0136 }         & \textrm{15.4656 }  & \textrm{14.9259 }  & \textrm{3.9754} & \textrm{-1.0327E-05 } & \textrm{End of optically thick mass transfer (U$_{2}$)}\\
\textrm{}       & \textrm{H}        & \textrm{48.1111 }  & \textrm{1.3618 }         & \textrm{24.9126 }  & \textrm{39.9391 }  & \textrm{4.0177} & \textrm{-8.8968E-08 } & \textrm{End mass transfer stage}\\
\textrm{}       & \textrm{I}        & \textrm{53.2305 }  & \textrm{1.3326 }		    & \textrm{ 0.3931 }  & \textrm{40.1799 }  & \textrm{4.7147} & \textrm{-3.5969E-09 } & \textrm{Helium depletion (${}^{4}$He)}\\
\hline
\textrm{Gainer} & \textrm{}         & \textrm{  }  		 & \textrm{ }         		& \textrm{  }  		 & \textrm{  }  	  & \textrm{} 		& \textrm{ }            & \textrm{}             \\
\textrm{}       & \textrm{a}        & \textrm{ 0.0000}  & \textrm{ 5.9000 }   		& \textrm{ 2.9088 }  & \textrm{ 2.5000}  & \textrm{ 4.3402} & \textrm{ -7.3593E-11} & \textrm{Zero Age Main Sequence (ZAMS)}             \\
\textrm{}       & \textrm{b}        & \textrm{ 0.1855}  & \textrm{ 5.9000 }   		& \textrm{ 2.6539 }  & \textrm{ 2.5000}  & \textrm{ 4.3230} & \textrm{ -1.1178E-10} & \textrm{Terminal Age Main Sequences (TAMS)}             \\
\textrm{}       & \textrm{U$_{1}$}  & \textrm{47.8906}  & \textrm{ 6.0897 }   		& \textrm{ 5.2028 }  & \textrm{ 2.5059}  & \textrm{ 4.3085} & \textrm{  5.3880E-05} & \textrm{Mass inversion}             \\
\textrm{}       & \textrm{X$_{2}$}  & \textrm{47.9236}  & \textrm{ 9.3392 }   		& \textrm{ 5.1596 }  & \textrm{ 2.3463}  & \textrm{ 4.4259} & \textrm{  9.6500E-05} & \textrm{Current stage}       \\
\textrm{}       & \textrm{c}        & \textrm{47.9272}  & \textrm{ 9.6612 }   		& \textrm{ 5.0893 }  & \textrm{ 2.5240}  & \textrm{ 4.4311} & \textrm{  8.4224E-05} & \textrm{End of mass accretion and relocation to the MS}             \\
\textrm{}       & \textrm{d}        & \textrm{47.9407}  & \textrm{10.1512 }  		& \textrm{ 4.6670 }  & \textrm{14.8279}  & \textrm{ 4.4301} & \textrm{  1.0703E-05} & \textrm{Second Main Sequence turnon}       \\
\textrm{}       & \textrm{e}        & \textrm{48.1111}  & \textrm{10.7985 }   		& \textrm{ 4.4084 }  & \textrm{39.9391}  & \textrm{ 4.4603} & \textrm{ -2.6812E-08} & \textrm{Exit from the threshold of rejuvenated B type stars}  \\
\textrm{}       & \textrm{f}        & \textrm{53.2305}  & \textrm{10.7917 }   		& \textrm{ 5.8267 }  & \textrm{40.1799}  & \textrm{ 4.4308} & \textrm{ -1.5193E-09} & \textrm{Stop}  \\
\hline
\end{tabular}
}
\end{center}
\label{tab:EVOL}
Notes: Exploring the evolutionary stages of DPV062: from the ZAMS to ${}^{4}$He depletion of the donor star. Detailed descriptions of key features are provided, along with corresponding ages measured in Mega years (Myr). RL means Roche lobe.
\end{table*}

We simulated the evolution of the binary system \var\, with \texttt{MESA} following \citet{2019MNRAS.483..862R}. The simulation initiates at the Zero-Age Main Sequence ($\mathrm{ZAMS}$) under a quasi-conservative mass transfer regime, which is maintained until the helium-4 mass fraction in the donor star's core ($\mathrm{X_{{}^4\mathrm{He}, c}}$) drops below $\mathrm{0.2}$. To identify the optimal model, we extensively explored the parameter space: the initial orbital period ($\mathrm{P_{i,o}}$) was varied from $\mathrm{1.0}$ to $\mathrm{6.0}$ days (steps of $\mathrm{0.1\ d}$); the initial stellar masses were explored with $\mathrm{M_{i, \text{d}}}$ ranging from $\mathrm{6.0}$ to $\mathrm{9.0\ M_{\odot}}$ and, conversely, $\mathrm{M_{i, \text{g}}}$ from $\mathrm{9.0}$ to $\mathrm{6.0\ M_{\odot}}$ (steps of $\mathrm{0.1\ M_{\odot}}$); and a metallicity of $\mathrm{Z = 0.006}$ was fixed, consistent with the assumed chemical composition of the Large Magellanic Cloud ($\mathrm{LMC}$). Key non-conservative mass loss parameters were optimized: the ${\alpha}$ parameter (fraction of mass lost from the donor as a fast Jeans-type wind) was explored from $\mathrm{9 \times 10^{-8}}$ to $\mathrm{9 \times 10^{-1}}$, and the ${\beta}$ parameter (fraction of mass ejected as a fast isotropic wind from the vicinity of the accretor after re-emission) was optimized, maintaining ${\beta >  \alpha}$. 
To independently modulate the envelope structure and evolutionary rate of both stellar components, we varied the mixing length parameter ($\mathrm{\alpha_{\text{ML}}}$), which governs the efficiency of convective energy transport in $\mathrm{MESA}$ via the Mixing Length Theory ($\mathrm{MLT}$). $\mathrm{\alpha_{\text{ML}}}$ is defined as the ratio between the mixing length and the star's pressure scale height. A high $\mathrm{\alpha_{\text{ML}}}$ (e.g., $\mathrm{\ge 1.0}$) implies efficient convection (near-adiabatic gradient), while a very low value simulates inefficient convection, significantly affecting the stellar radius and evolution. As a reference,  $\mathrm{\alpha_{\text{ML}}}$ varies systematically in the range of 1.7 - 2.4 for stars of solar metallicity \citep{2015A&A...573A..89M}. However, uncertainties remain for stars quite different from the Sun (e.g., low metallicity, evolved giants, very low or high mass),  limiting the predictive reliability of 1D evolutionary models in those regimes.
In this study, we explored the $\mathrm{\alpha_{\text{ML}}}$ value from $\mathrm{0.0001}$ to $\mathrm{4.0}$ for the donor star, keeping the value for the gainer at 1.8. 

The determination of the best-fitting initial parameters was achieved through a rigorous minimization analysis using the
chi-squared ($\chi^2$) statistic. This method compares the theoretical outputs from the $\texttt{MESA}$ simulations with the observed physical parameters. To refine the fitting process, a weighted $\chi^2$ function was implemented, where individual contributions from the observed parameters were scaled using weight factors (${w_i}$), prioritizing certain observables (like masses or orbital period) over others (like luminosities or temperatures).

The best model, which yielded the lowest chi-squared value of $\chi^2 = 0.1486$, converged with the following initial conditions: donor mass $\mathrm{M_{i,d} = 6.3\,M_{\odot}}$, accretor mass $\mathrm{M_{i,g} = 5.9\, M_{\odot}}$, and initial orbital period $\mathrm{P_{i,o} = 2.5\,d}$. The corresponding mass transfer parameters were $\mathrm{\alpha = 8 \times 10^{-5}}$ $M_{\odot}$/yr
and $\mathrm{\beta = 1 \times 10^{-4}}$ (${\delta}$ mode was set to $\mathrm{0}$).
The donor ${\alpha_{\text{ML}}}$ parameter required for convergence was ${4.0}$. This value is high for the standards, and could be influenced by the fact that the upper levels of the donor atmosphere are being affected by the mass transfer. We notice that lower values fail in reproducing well the position of the donor in the luminosity-temperature diagram. The best fit model at present age (47.92 Myr) is characterized by
$\mathrm{M_{g} = 9.34\, M_{\odot}}$, $\mathrm{M_{d} = 2.83\, M_{\odot}}$,  $\mathrm{R_{g} = 5.16\, R_{\odot}}$, $\mathrm{R_{d} = 10.21\, R_{\odot}}$, $\mathrm{T_{g} = 26662\,K}$ 
and $\mathrm{M_{d} = 6973\,K}$.

The evolutionary tracks of the gainer and donor are shown in Fig. \ref{fig:LT}, with labels indicating the evolutionary stages described in Table \ref{tab:EVOL}.
Variations in the central helium and hydrogen abundances reflect hydrogen depletion and helium enrichment in the cores of both stars. However, at the current evolutionary stage, the gainer exhibits a rejuvenation effect, resulting from the increase in its central hydrogen content supplied by the donor during mass transfer, with the accreted hydrogen being efficiently mixed from the surface into the stellar interior (see Fig. \ref{fig:LT}).

\section{Discussion}

Before we proceed with the interpretation of our results, it's important to be aware of the limitations of our model. A key simplification is our focus exclusively on circumstellar material within the disk, thereby overlooking any light contributions originating from regions above or below the disk plane. Additionally,  we have assumed the gainer is surrounded by an accretion disk, disregarding other potential light sources such as jets, winds, and outflows. Nonetheless, the model allows for the possibility of accounting for variations in disk emissivity that are dependent on azimuthal position through the inclusion of two disk spots. Despite these constraints, and drawing on prior research on algols with disks, our model likely captures the principal light sources for the continuum light.  This is evidenced by the close correspondence between the model light curve and the orbital and long-term light curves over the span of  32.3 years of observation (the time covered by the $I$-band OGLE time series). This finding that the long cycle of \var\, is a consequence of its variable accretion disk is consistent with recent claims that DPV cycles represent disk's cyclic structural changes \citep[e.g.][]{2018MNRAS.477L..11G, 2025A&A...701A..90G}.  In this Section, we discuss the evolutionary stage of the binary and also we provide a test for the magnetic dynamo hypothesis of the long-term cycle. 

\subsection{On the disk size and mass transfer rate}

We run evolutionary models and find the best one reproducing the present state of the system. Although the match is not perfect, we can certainly infer that the system is product of the evolution of a stellar pair that experiences strong mass transfer in a semidetached stage with Roche lobe overflowing. 

The fractional radius of the gainer is $R_1/a$ = 0.143. This value reveals a tangential impact system, where the gas stream hits the gainer almost tangentially, provoking the formation of an accretion disk \citep{1975ApJ...198..383L}. 

In principle, the disk can be stable until the last non-intersecting orbit defined by the tidal radius \citep[][Eq. 2.61]{1977ApJ...216..822P, 1995CAS....28.....W}:

\begin{eqnarray}
\frac{R_{\rm{t}}}{a_{\rm{orb}}}= \frac{0.6}{1 + q},
\end{eqnarray}

\noindent we get $R_t/a_{orb}$ = 0.476, or $R_t$ =  17.10 $R_{\odot}$.  We observe that, during all the observing epochs (except set 6 when is 17.8) the disk outer radius keeps inside the volume defined by the tidal radius. This discards possible influences of tidal forces on shaping the outer parts of the disk.

The value of the mass transfer rate of 9.7 $\times$ 10$^{-5}$ \msun/yr, must be  consistent with the constancy of the orbital period. For a binary system ongoing conservative mass transfer, we should expect a change of the orbital period of \citep{1963ApJ...138..471H}:

\begin{equation}
\frac{\dot{P_{\rm o}}}{P_{\rm o}} = 3\dot{M_2}(\frac{1}{M_2}-\frac{1}{M_1}) = 7.8 \times 10^{-5} \rm{yr^{-1}}.
\end{equation}

\noindent This means a change of 5.42 $\times$ 10$^{-4}$ days per year, or 0.01750 days during the whole observing window. This is almost 45 times larger than the period error of 0.000393 days. As not observed, the actual mass transfer rate is much smaller than 9.7 $\times$ 10$^{-5}$ \msun/yr or systemic mass loss is important for the system angular momentum balance. The mass transfer rate compatible with the constancy of the orbital period should be lower than 2.2 $\times$ 10$^{-6}$  $M_{\odot}$/yr. In case of important systemic mass loss in the past, the system total mass should be smaller than calculated.  

In principle, mass loss in a semidetached binary undergoing Roche lobe overflow can occur through the outer Lagrangian points L$_{2}$ and L$_{3}$, via stellar winds, or even through a jet of material directed roughly perpendicular to the orbital plane. These mechanisms are either theoretically predicted \citep{2025ApJ...990..172S} or have been observationally reported in some DPVs \citep{2012MNRAS.427..607M, 1996A&A...312..879H}. However, the evolutionary pathways of Roche-lobe overflowing binaries remain highly uncertain, as key processes such as the efficiency of accretion onto the gainer and the resulting spin-up are still poorly constrained. Consequently, modeling non-conservative evolution during episodes of systemic mass loss remains a major theoretical challenge \citep{2024ARA&A..62...21M} although recent results indicate predominantly conservative cases in a subset of 16 Be+sdOB binaries \citep{2025ApJ...990L..51L}.

\begin{figure}
\scalebox{1}[1]{\includegraphics[angle=0,width=8cm]{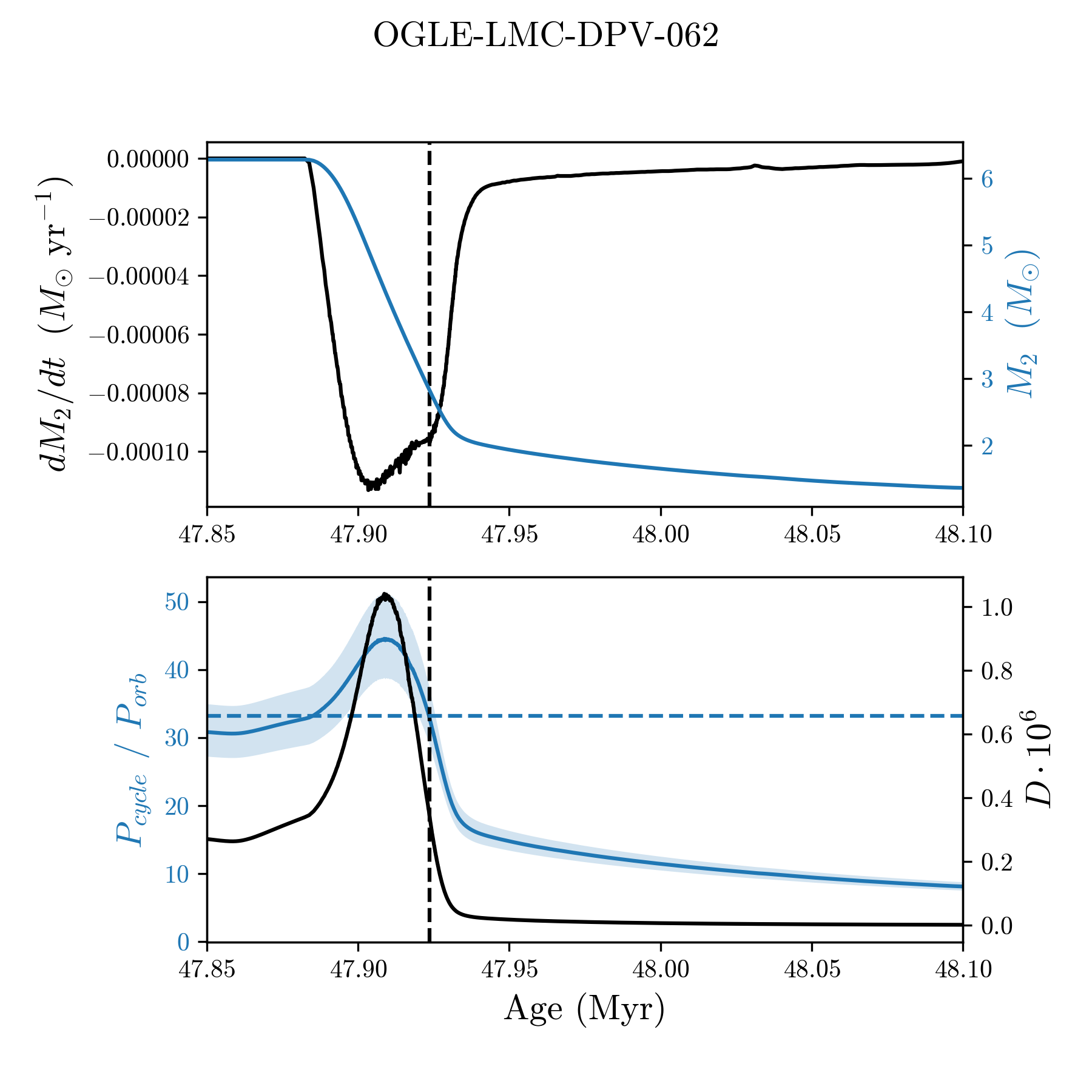}}
\caption{Top: mass-transfer rate and evolution of the donor star mass.  Bottom: evolution of the theoretical value of $P_{\mathrm{cycle}}/P_{\mathrm{orb}}$ considering $\gamma = 0.27 \pm 0.01$ (blue region), together with the estimated dynamo number.  The horizontal line shows the observed relation $P_{\mathrm{long}}/P_{\mathrm{orb}} = 33.2$. In both panels, the vertical line indicates the age, according to the best \texttt{MESA} evolutionary model, at which the system reaches an orbital period of $P_{\mathrm{orb}} = 6\fd904858$.}
\label{fig:dynamo}
\end{figure}

\subsection{Test for the magnetic dynamo model}

Assuming that the magnetic dynamo mechanism acting on the donor star is the origin of the long cycle, it is possible to estimate the theoretical duration of this cycle from the relations between the Keplerian period ($P_{\mathrm{Kep}}$), the convective velocity ($v_{c}$), and the Rossby number ($R_{o}$). The latter can be evaluated, following \citet{2017A&A...602A.109S}, as:

\[
R_{o} = 11.5 \, \dfrac{v_{c} \, P_{\mathrm{Kep}}}{R_{2}} \, \dfrac{R_{\odot}}{\mathrm{km\,s^{-1}\,yr}}.
\]

where $R_{2}$ is the radius of the donor star. In addition, the Rossby number is linked to the dynamo number ($D$) through $D = R_{o}^{-2}$, so that the duration of the magnetic cycle acting on the donor can be estimated from \citep[][and references therein]{2017A&A...602A.109S}:

\[
P_{\mathrm{cycle}} = D^{\gamma} \times P_{\mathrm{orb}},
\]

\noindent where $\gamma$ has typical values between 1/3 and 5/6 and the equation shows the link between the rotational period of a single star (assumed here equal to the orbital period) and the cycle length of its magnetic dynamo \citep{1993ApJ...414L..33S,1996ApJ...460..848B}. 

Following this approach, and using the physical parameters of the donor obtained from the best evolutionary model computed with \texttt{MESA}, we can trace the temporal evolution of the ratio $P_{\mathrm{cycle}}/P_{\mathrm{orb}}$ as the radius, luminosity, and stellar mass of the donor change during the Roche Lobe Overflow (RLOF) phase.

In Fig.\,\ref{fig:dynamo} we note a good agreement between the theoretical value of $P_{\mathrm{cycle}}/P_{\mathrm{orb}}$ and the observed ratio $P_{\mathrm{long}}/P_{\mathrm{orb}}$. This fit is achieved by adopting a value of $\gamma = 0.27 \pm 0.01$, slightly below than $\gamma = 0.31$ predicted for DPVs, but within the expected range for active stars  \citep[$\sim 0.25{-}0.83$;][]{2017A&A...602A.109S}. 

We confirm a pronounced variation in the donor's dynamo number during the mass-transfer episode as reported by \citet{2019BAAA...61..107S}. These variations are linked to structural changes in the donor star, particularly in its radius and rotational velocity. It is therefore plausible that the conditions required for a magnetic dynamo operate only within a limited time window during the overall mass-transfer phase, a period compatible with the observed DPV $\gamma$ values. However, the observation that the long-term cycle could change or even disappears in very short time scales, as occurs in AU\,Mon \citep{2025A&A...700A.217C}, indicates that other mechanisms beside changes in the internal donor structure influence the long cycles.

\subsection{Comparison with others DPVs} 

\var\, shows stellar parameters in the range exhibited by others DPVs. Regarding the long-cycle it shows a behavior similar to OGLE-BLG-ECL-157529, OGLE-LMC-ECL-14413, and RX~Cas;  
an increase in the mass transfer rate produces a thicker disk and a fainter system \citep{2021A&A...653A..89M, 2022A&A...666A..51M, 2025A&A...693A.217M}. However, in OGLE-LMC-DPV-097 the disk radius is the primary factor governing the brightness at both maximum and minimum light \citep{2018MNRAS.477L..11G}, whereas in OGLE-LMC-DPV-065, showing double-hump long-cycle light curve, the disk central thickness decreases at maximum brightness, and both the disk temperature and radius display complex cyclic variations \citep{2025A&A...698A..56M}. It is clear that the behaviour is not necessarily the same for all DPVs, and that more research is needed to understand how the disk reacts to changes in the mass transfer rate.

\section{Conclusion}

We have analyzed the remarkable long-term light curve of the DPV system \var, covering a timespan of 32.3 years and obtaining, for the first time, a physical description of the whole system. Our model successfully reproduces the overall $I$-band variability, accounting for both the orbital modulation with a period of 6\fd90 and the long-term DPV cycle of 229\fd7.

We find the system consisting of a B-type main-sequence star, approximately of spectral type B1.5, accreting matter from a cooler A9-F0 type giant, according to the classification scheme of \citet{1988BAICz..39..329H}. Our study traces the system's evolutionary path, indicating that it is currently undergoing an episode of rapid mass transfer as the donor fills its Roche lobe. We have estimated fundamental stellar parameters for the binary. The components have masses of 10.36 $\pm$ 1.14 \msun\, and 2.73 $\pm$  0.30 \msun, radii of 5.14 $\pm$  0.98 \rsun\, and 9.71 $\pm$    0.56  \rsun, and effective temperatures of 24400 $\pm$   2400 K and  7100 $\pm$   700 K, respectively. The orbital separation is 35.9 $\pm$  1.4 \rsun, and the surface gravities  are log g = 4.03 $\pm$  0.17 and 2.90 $\pm$   0.07. These estimates should be treated with caution, since the possibility of non-conservative mass transfer, along with related uncertainties not considered here, cannot be excluded.

Our model incorporates the contribution of an accretion disk with an average radius of 14.5 \rsun\, and includes two hot shock regions at the disk's outer edge. 
We detect substantial variations in the disk's vertical structure, suggesting that turbulence and large scale motions may play a role in sustaining its thickness. 

We find that the normalized  mass-transfer rate varies in phase with the long cycle, reaching at maximum 16 times its minimum value, when the system's total brightness is at its minimum. At this stage, the disk attains its greatest vertical thickness at the inner edge, almost 5 \rsun, obscuring a larger portion of the gainer. In our model, the disk's variability is primarily expressed through changes in vertical extension, while its radius and the temperature at the outer rim exhibit only minor variations throughout the long cycle. Additionally, hot and bright spots are present on the disk surface at roughly constant azimuths, located opposite the donor star.

Although no direct evidence for magnetism is found, a good match is observed between the prediction of the magnetic dynamo model of \citet{2017A&A...602A.109S} and the observed ratio between the orbital and long-term cycle periods.

\begin{acknowledgements}
 We thank the anonymous referee for the useful comments and suggestions on the first version of this manuscript. 
We acknowledge support by the ANID BASAL project Centro de Astrof{\'{i}}sica y Tecnolog{\'{i}}as Afines ACE210002 (CATA). This work has been co-funded by the National Science Centre, Poland, grant No. 2022/45/B/ST9/00243.
\end{acknowledgements}

{}

 \begin{appendix}
 
 \section{Error estimates}

We estimate the errors of mass, radius and orbital separation assuming an uncertainty in the temperature and using an analytical approach. The luminosity follows the Stefan-Boltzmann law:

\begin{equation}
L = 4 \pi R^2 \sigma T^4
\end{equation}

\noindent
where R is the radius of the star, $\sigma$ is the Stefan-Boltzmann constant, and $T$ is the effective temperature. Now, an error in the temperature, 
$\Delta T$, affects the luminosity due to the dependence $L$ $\propto$ $T^{4}$. If we make an error $\Delta T$ in measuring the effective temperature of the star, the relative error in the luminosity will be:

\begin{equation}
\frac{\Delta L}{L} = 4 \frac{\Delta T}{T} 
\end{equation}

Given that for a main sequence star $L \propto M^{\alpha}$, an error in the luminosity, $\Delta L$, translates into an error in the mass, $\Delta M$, as follows:

\begin{equation}
\frac{\Delta M}{M} = \frac{1}{\alpha} \frac{\Delta L}{L} = \frac{4}{\alpha} \frac{\Delta T}{T}
\end{equation}

If we use $\alpha \approx$ 3.5 for a main sequence star, and consider that for a giant star the exponent should be larger because of the increasing radius, the relative error in the mass of the donor would be:

\begin{equation}
\frac{\Delta M}{M} < \frac{4}{3.5} \frac{\Delta T}{T} \approx  1.14 \frac{\Delta T}{T} 
\end{equation}

On the other hand, the orbital separation in a binary system is given by Kepler's third law:

\begin{equation}
a = (\frac{G(M_1+M_2)P^2}{4\pi^2})^{1/3}
\end{equation}

An error in the stellar masses affects the orbital separation as follows:

\begin{equation}
\frac{\Delta a}{a} = \frac{1}{3} \frac{\Delta (M_1 + M_2)}{M_1 + M_2} 
\end{equation}

Assuming the relative error is similar for both stars, we can write:

\begin{equation}
\frac{\Delta a}{a} = \frac{1}{3} \frac{4}{\alpha} \frac{\Delta T}{T} 
\end{equation}

For $\alpha$ $>$ 3.5:

\begin{equation}
\frac{\Delta a}{a} <  \frac{1.14}{3} \frac{\Delta T}{T}  \approx 0.38    \frac{\Delta T}{T}
\end{equation}

From the above equations, and assuming a temperature uncertainty of 10\%, we get masses with uncertainty of less than 11\% and orbital separation with uncertainty of less than 4\%.

To calculate the error in the radius of the giant star filling its Roche lobe, we need to consider how the errors in the mass ratio (\(q = M_1 / M_2\)) and the orbital separation (\(a\)) affect the error in the Roche lobe radius (\(R_L\)).

Recall that the Roche lobe radius is given by the empirical Eggleton formula:

\begin{equation}
\frac{R_L}{a} = \frac{0.49 q^{2/3}}{0.6 q^{2/3} + \ln(1 + q^{1/3})}
\end{equation}

To find the error in \(R_L\) (the Roche lobe radius, which will approximately be the radius of the giant star), we applied the formula for error propagation. This allows us to calculate the error in \(R_L\) as a function of the errors in \(q\) and \(a\).

Since \(R_L\) depends on two variables, \(q\) and \(a\), the error in \(R_L\), denoted as \(\Delta R_L\), can be computed using the following error propagation formula:

\begin{equation}
\frac{\Delta R_L}{R_L} = \sqrt{\left( \frac{\partial R_L}{\partial q} \cdot \frac{\Delta q}{q} \right)^2 + \left( \frac{\partial R_L}{\partial a} \cdot \frac{\Delta a}{a} \right)^2}
\end{equation}

First we calculate the partial derivative with respect to \(q\):

\tiny
   \begin{equation}
   \frac{\partial}{\partial q} \left( \frac{R_L}{a} \right) = \frac{0.49 \cdot \frac{2}{3} q^{-1/3}}{\left(0.6 q^{2/3} + \ln(1 + q^{1/3}) \right)} - \frac{0.49 q^{2/3} \cdot \left( 0.6 \cdot \frac{2}{3} q^{-1/3} + \frac{1}{3(1 + q^{1/3}) q^{2/3}} \right)}{\left(0.6 q^{2/3} + \ln(1 + q^{1/3})\right)^2}
   \end{equation}
\normalsize

Then,  we calculate the partial derivative with respect to \(a\). The dependence of \(R_L\) on \(a\) is straightforward, as \(R_L \propto a\). Thus:

   \begin{equation}
   \frac{\partial R_L}{\partial a} = \frac{R_L}{a}
   \end{equation}
   
For $q$ = 0.26 with an associated error of  0.04, and assuming a 4\% relative error in the orbital separation, we obtain a relative error of 4.1\% for the Roche Lobe radius, which also corresponds to the donor star's radius.

 If the gainer follows a mass-radius relationship $M \propto R^{\beta}$ with $\beta$ between 0.57 and 0.7, we get the fractional error in the gainer radius less than 1.75  times the fractional error in the gainer mass. If this last is  11\%, then the gainer fractional radius uncertainty is less than 19\%.

Finally, the surface gravity in solar units is given by:

\begin{equation}
 \log g = \log \left( \frac{M}{R^2} \right) + \log g_{\odot}
   \end{equation}

\noindent
where $\log g_{\odot}$ is the solar surface gravity. The error is given by:

\begin{equation}
\Delta (\log g) = \sqrt{\left( \frac{\Delta M}{M \ln(10)} \right)^2 + \left( \frac{2 \Delta R}{R \ln(10)} \right)^2}
 \end{equation}

Using the fractional errors quoted above, we derive the errors for log\,g given in Table \ref{tab:system}, viz. 4\% por the gainer and 2\% for the donor.

\clearpage 
\section{Additional material}

\begin{table*}[h]
\centering
\caption{ Coefficients for the fits shown in Fig.\ref{fig:LCresults}.}
\label{tab:fits}
\begin{tabular}{crrrrr}
\hline 
\hline
Parameter & Coefficient &Value &Standard Error &t-Value & Prob $>$ |t|\\
\hline
$\dot{M}$&	Intercept&	2.89&	4.14	   &      0.698&0.497\\
$\dot{M}$&	B1&	39.07	     &   115.43&	0.338	&0.740\\
$\dot{M}$&	B2&	-791.74&	994.28&	-0.796	&0.440\\
$\dot{M}$&	B3&	4451.85&	3706.56&	1.201	&0.251\\
$\dot{M}$&	B4&	-9513.44&	6716.33&	-1.416	&0.180\\
$\dot{M}$&	B5&	8628.78&	5824.48&	1.481	&0.162\\
$\dot{M}$&	B6&	-2810.27&	1935.34&	-1.452	&0.170\\
I$_{0.25}$&	Intercept&	15.72&	0.01&	1760.052&0.000   \\
I$_{0.25}$&	B1&	-0.59	       &  0.25	     &    -2.384&0.033\\
I$_{0.25}$&	B2&	10.57	       &    2.14&	4.934	  & 0.000\\
I$_{0.25}$&	B3&	-36.99       &    	7.99&	-4.631	  & 0.000\\
I$_{0.25}$&	B4&	59.92	       &   14.47&	4.139	  & 0.001\\
I$_{0.25}$&	B5&	-48.63	       &   12.55&	-3.874	  & 0.002\\
I$_{0.25}$&	B6&	15.72	       &    4.17&	3.770	   &0.002\\
d$_{\rm c}$&	Intercept&	2.48	&1.02	      &  2.421	&0.031\\
d$_{\rm c}$&	B1&	37.79	        & 28.54	&1.324	&0.208\\
d$_{\rm c}$&	B2&	-252.66	&245.80	&-1.028	&0.323\\
d$_{\rm c}$&	B3&	693.97	&916.36	&0.757	&0.462\\
d$_{\rm c}$&	B4&	-977.09	&1660.45&	-0.588	&0.566\\
d$_{\rm c}$&	B5&	720.74	&1439.96	&0.500	&0.625\\
d$_{\rm c}$&	B6&	-222.50	&478.47	&-0.465	&0.650\\
\hline
\end{tabular}
\end{table*}

\begin{figure*}
\scalebox{1}[1]{\includegraphics[angle=0,width=18cm]{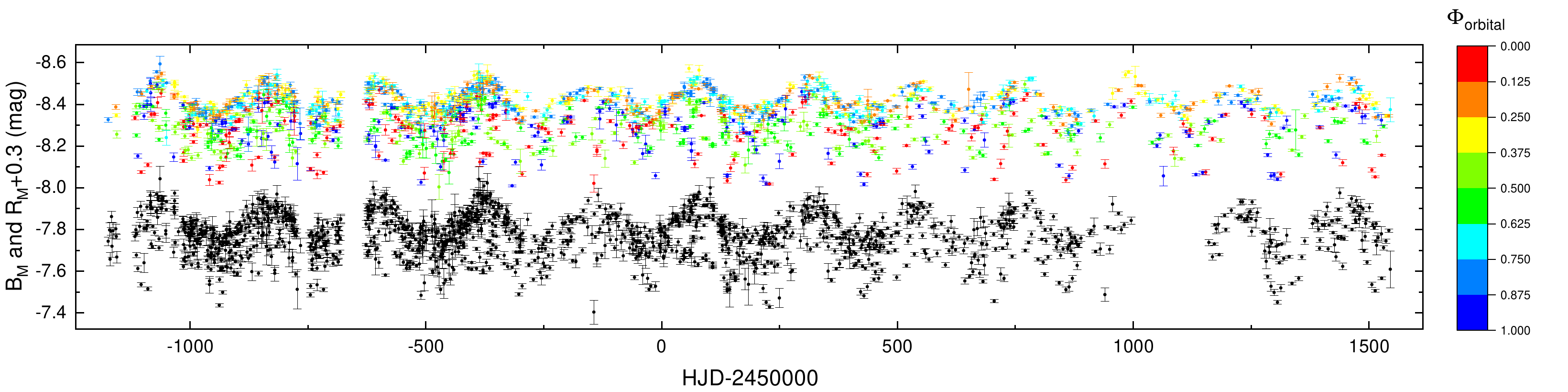}}
\caption{MACHO $B_M$ (colored) and $R_M$  light curves.}
\label{fig:machoLCs}
\end{figure*}

\begin{figure*}
\scalebox{1}[1]{\includegraphics[angle=0,width=18cm]{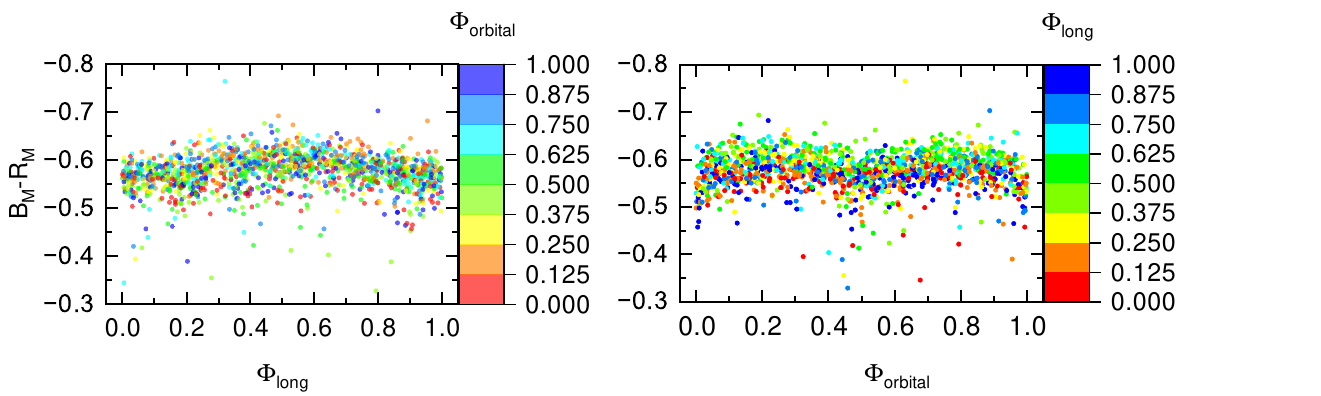}}
\caption{MACHO $B_M$-$R_M$  color versus orbital and long-term cycle phases.}
\label{fig:machocolor}
\end{figure*}

\begin{figure*}
\scalebox{1}[1]{\includegraphics[angle=0,width=18cm]{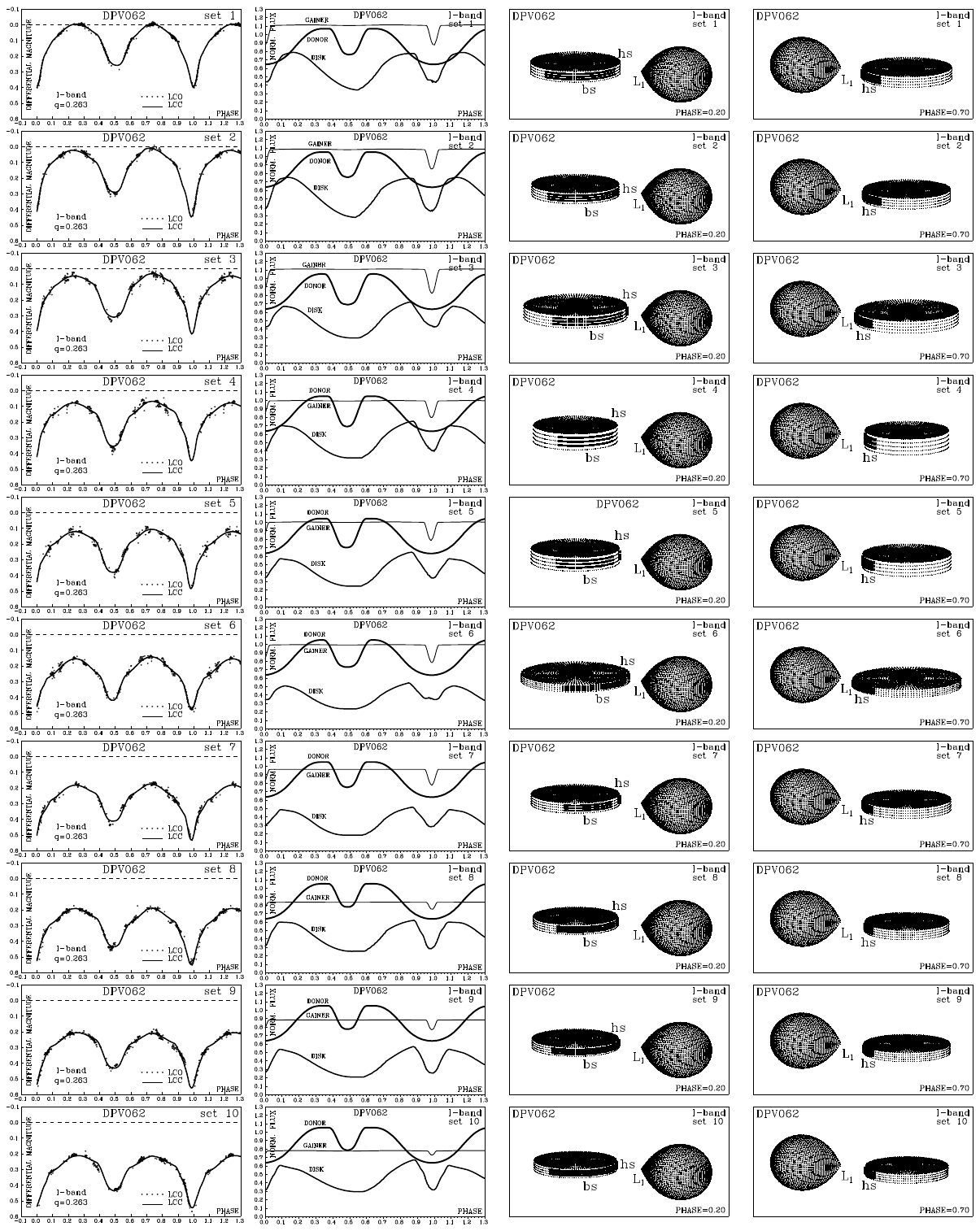}}
\caption{Orbital light curve fits for individual datasets 1-10.}
\label{fig:LC1}
\end{figure*}

\begin{figure*}
\scalebox{1}[1]{\includegraphics[angle=0,width=18cm]{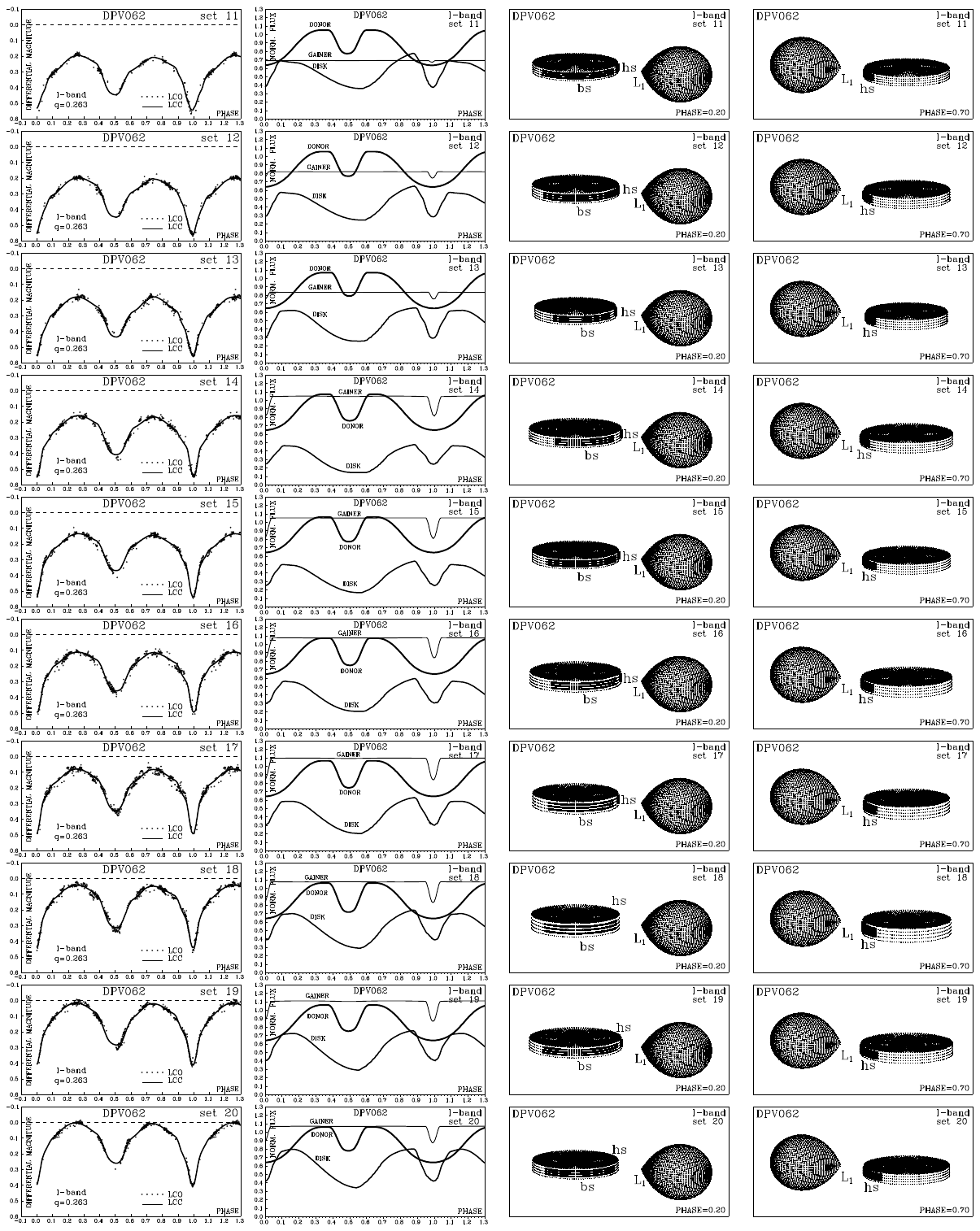}}
\caption{Orbital light curve fits for individual datasets 11-20.}
\label{fig:LC2}
\end{figure*}

\begin{figure*}
\scalebox{1}[1]{\includegraphics[angle=0,width=8.5cm]{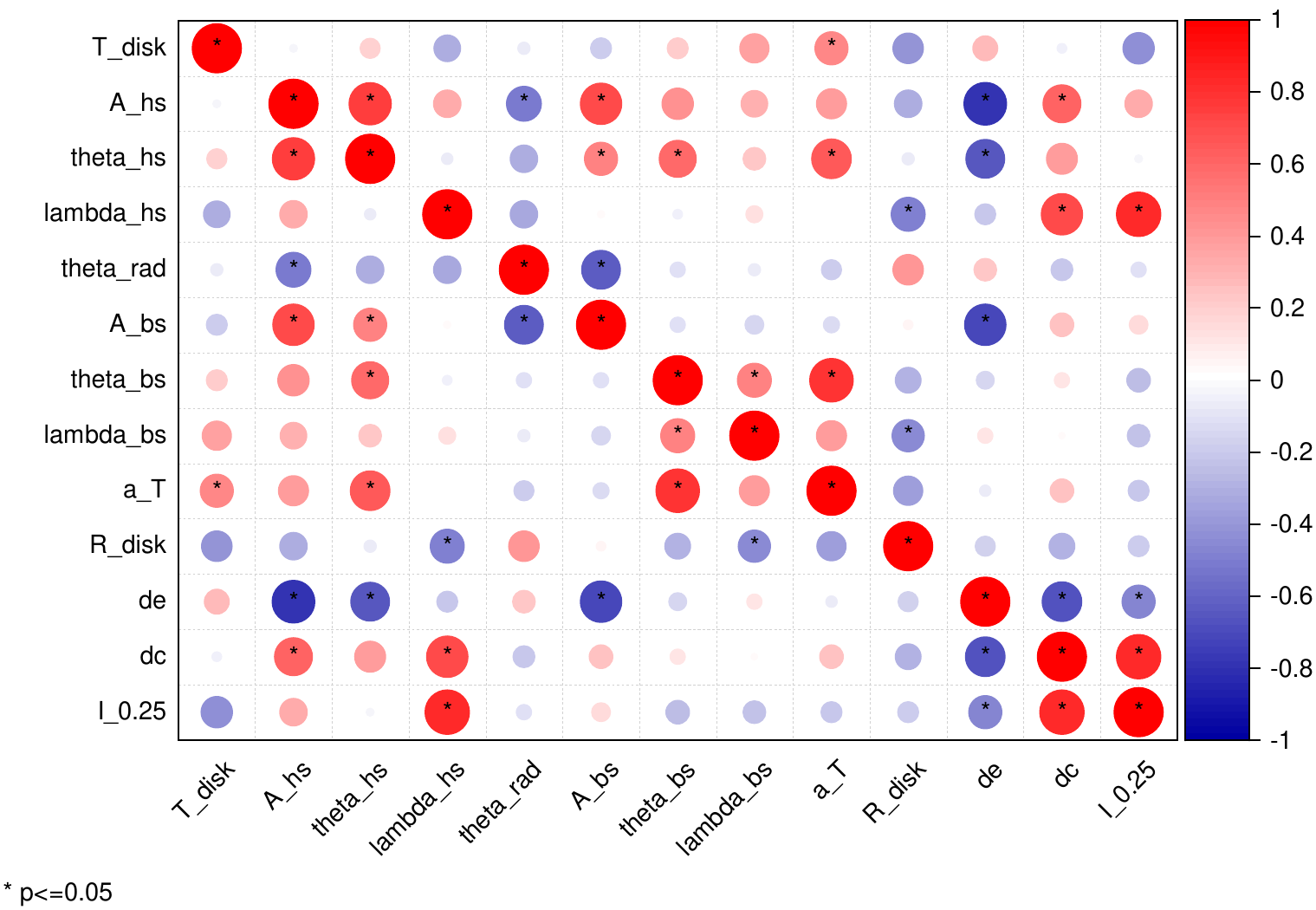}}
\caption{Correlation matrix for the fit's parameters showing with asterisks those cases with p-value equal or under 0.05. The right vertical color scale shows the correlation coefficient. }
\label{fig:SigMatrix}
\end{figure*}

\end{appendix}
\end{document}